\documentclass[aps,prd,reprint,superscriptaddress,showpacs,showkeys,preprintnumbers,floatfix,a4paper,nofootinbib]{revtex4-1}
\usepackage{color}
\usepackage{amsmath}
\usepackage{amssymb}
\usepackage{dcolumn}  
\usepackage{slashed}
\usepackage{graphicx}
\usepackage{mathtools}
\usepackage[utf8x]{inputenc}

\newcommand{\GX}[1]{G_{\rm{#1}}}
\newcommand{\zv}{Z_{\rm V}}
\newcommand{\bv}{b_{\rm V}}
\newcommand{\cv}{c_{\rm V}}

\newcommand{\ripm}{\text{RI$'$\kern-0.06667em/MOM}}

\newcommand{\mq}{m_{\text{q}}}

\if@mathematic
\renewcommand{\vec}[1]{\ensuremath{\mathchoice
    {\mbox{\boldmath$\displaystyle\mathbf{#1}$}}
    {\mbox{\boldmath$\textstyle\mathbf{#1}$}}
    {\mbox{\boldmath$\scriptstyle\mathbf{#1}$}}
    {\mbox{\boldmath$\scriptscriptstyle\mathbf{#1}$}}}}
\else
\renewcommand{\vec}[1]{\ensuremath{\mathchoice
    {\mbox{\boldmath$\displaystyle#1$}}
    {\mbox{\boldmath$\textstyle#1$}}
    {\mbox{\boldmath$\scriptstyle#1$}}
    {\mbox{\boldmath$\scriptscriptstyle#1$}}}}
\fi

\newcommand{\eVdist}{\kern-0.06667em}

\newcommand{\Gev}{{\text{Ge}\eVdist\text{V\/}}}

\newcommand{\rme}{{\mathrm{e}}}
\newcommand{\fm}{{\mathrm{fm}}}
\newcommand{\psibar}{{\overline{\psi}}}
\newcommand{\rb}[1]{\raisebox{1.5ex}[-1.5ex]{#1}}

\begin{document}
\allowdisplaybreaks

\preprint{MITP/15-026}
\preprint{HIM-2015-01}
\preprint{CP3-Origins-2015-012}
\preprint{DIAS-2015-12}

\title{Nucleon electromagnetic form factors in two-flavour QCD}



\author{S.~Capitani}
\affiliation{PRISMA Cluster of Excellence and Institut f\"ur Kernphysik, University of Mainz, Becher-Weg 45, 55099 Mainz, Germany}
\affiliation{Helmholtz Institute Mainz, University of Mainz, 55099 Mainz, Germany}
\author{M.~Della~Morte}
\affiliation{CP3-Origins, University of Southern Denmark, Campusvej 55, 5230 Odense M, Denmark}
\author{D.~Djukanovic}
\affiliation{Helmholtz Institute Mainz, University of Mainz, 55099 Mainz, Germany}
\author{G.~von~Hippel}
\affiliation{PRISMA Cluster of Excellence and Institut f\"ur Kernphysik, University of Mainz, Becher-Weg 45, 55099 Mainz, Germany}
\author{J.~Hua}
\affiliation{PRISMA Cluster of Excellence and Institut f\"ur Kernphysik, University of Mainz, Becher-Weg 45, 55099 Mainz, Germany}
\author{B.~J\"ager}
\affiliation{Department of Physics, College of Science, Swansea University, Swansea~SA2~8PP, United Kingdom}
\author{B.~Knippschild}
\thanks{\textit{present address:} Helmholtz-Institut f\"ur Strahlen- und Kernphysik (Theorie), Universit\"at Bonn, 53115 Bonn, Germany}
\affiliation{PRISMA Cluster of Excellence and Institut f\"ur Kernphysik, University of Mainz, Becher-Weg 45, 55099 Mainz, Germany}
\author{H.B.~Meyer}
\affiliation{PRISMA Cluster of Excellence and Institut f\"ur Kernphysik, University of Mainz, Becher-Weg 45, 55099 Mainz, Germany}
\affiliation{Helmholtz Institute Mainz, University of Mainz, 55099 Mainz, Germany}
\author{T.D.~Rae}
\email{thrae@uni-mainz.de}
\thanks{\textit{present address:} Bergische Universit\"at Wuppertal, Gaussstr. 20, D-42119~Wuppertal, Germany}
\affiliation{PRISMA Cluster of Excellence and Institut f\"ur Kernphysik, University of Mainz, Becher-Weg 45, 55099 Mainz, Germany}
\author{H.~Wittig}
\email{hartmut.wittig@uni-mainz.de}
\affiliation{PRISMA Cluster of Excellence and Institut f\"ur Kernphysik, University of Mainz, Becher-Weg 45, 55099 Mainz, Germany}
\affiliation{Helmholtz Institute Mainz, University of Mainz, 55099 Mainz, Germany}

\noaffiliation

\date{\today}


\begin{abstract}
We present results for the nucleon electromagnetic form factors,
including the momentum transfer dependence and derived quantities
(charge radii and magnetic moment).  The analysis is performed using
$\mathcal{O}(a)$ improved Wilson fermions in $N_f=2$ QCD measured on
the CLS ensembles. Particular focus is placed on a systematic
evaluation of the influence of excited states in three-point
correlation functions, which lead to a biased evaluation, if not
accounted for correctly. We argue that the use of summed operator
insertions and fit ans\"atze including excited states allow us to
suppress and control this effect. We employ a novel method to perform
joint chiral and continuum extrapolations, by fitting the form factors
directly to the expressions of covariant baryonic chiral effective
field theory. The final results for the charge radii and magnetic
moment from our lattice calculations include, for the first time, a
full error budget. We find that our estimates are compatible with
experimental results within their overall uncertainties.
\end{abstract}



\maketitle


\section{Introduction\label{sec_intro}}

The electromagnetic form factors, $\GX{E}$ and $\GX{M}$, of the
nucleon encode information on the distribution of charge and
magnetization and are among the key quantities describing its internal
structure. Experimental measurements of these quantities in $ep$
scattering processes have a long history (see, e.g. the review
\cite{Perdrisat:2006hj}) and have been pushed to ever higher
precision\,\cite{Bernauer:2010wm,Bernauer:2013tpr}. In spite of the
fact that nucleon electromagnetic form factors have been studied
extensively in theory and experiment, there are several open
questions. The first concerns the deviation between the ratio
$\GX{E}/\GX{M}$ as determined using the traditional Rosenbluth
separation technique and the result obtained from recoil
polarization\,\cite{Jones:1999rz,Gayou:2001qd,Punjabi:2005wq,Puckett:2010ac}
at squared momentum transfers $Q^2$ larger than $1~{\rm
  GeV}^2$. Secondly, prompted by the observed discrepancy between the
proton charge radius extracted from the Lamb shift in muonic
hydrogen\,\cite{Pohl:2010zza,Antognini:1900ns} and the value obtained
by using the electron as a probe\,\cite{Bernauer:2013tpr,Mohr:2012tt},
there is a strong interest in new experimental measurements of form
factors in the regime of very small $Q^2$, as well as in further
theoretical studies, in order to reduce the inherent systematics. The
third open issue concerns our understanding of the internal structure
of the nucleon in terms of the underlying gauge theory of QCD. Nucleon
form factors have been studied extensively in simulations of QCD on a
space-time lattice
\cite{Alexandrou:2006ru,nuclFF:RBC08_nf2,Yamazaki:2009zq,
  Syritsyn:2009mx,nuclFF:QCDSF_lat09,Bratt:2010jn,nuclFF:QCDSF_lat10,
  Alexandrou:2011db,Gockeler:2011ze,Collins:2011mk,Green:2013hja,
  Jager:2013kha,Green:2014xba}, and although these calculations are
quite straightforward, they mostly fail in reproducing the
experimentally observed $Q^2$-dependence of $\GX{E}$ and $\GX{M}$. As
a consequence, lattice estimates for the electric charge radius
derived from the slope of $\GX{E}$ at vanishing $Q^2$ are typically
underestimated compared to the results derived from $ep$ scattering
data. It is widely believed that systematic errors in lattice
calculations must be held responsible for this deviation.

In addition to systematic errors induced by non-zero lattice spacings, finite
volumes and by uncertainties associated with the chiral extrapolation, the
issue of contamination from excited states in calculations of nucleon
correlation functions has recently come to the fore as a possible explanation
for the deviation between experimental and lattice estimates of the electric
charge radius.

In this paper we present a detailed investigation of systematic effects in
lattice calculations of nucleon form factors arising from excited state
contributions. In particular, we apply the technique of summed operator
insertions \cite{smear:Gaussian89,Maiani:1987by,Doi:2009sq} which has proved
very useful in our earlier calculation of the axial charge of the
nucleon\,\cite{Capitani:2012gj}. Furthermore, we address in detail the chiral
extrapolation to the physical pion mass, by employing several variants of
baryonic Chiral Perturbation Theory (ChPT).

Our simulations are performed in two-flavour QCD with a
mass-degenerate doublet of up- and down-quarks. Since excited state
contamination is an issue for lattice simulations with any number of
dynamical quarks, the question whether estimates for nucleon charge
radii and magnetic moments may be biased can be adequately addressed
in this set-up. There is ample evidence\,\cite{Colangelo:2010et} that
there are no discernible differences between QCD with $N_{\rm{f}}=2$
and $N_{\rm{f}}=2+1$ flavours at the few-percent level. Therefore, the
observed deviation between lattice QCD and experiment is far too large
to be explained by the presence or absence of a dynamical strange
quark.

Our central findings include the observation that excited-state
contaminations have a sizeable influence on the form factors extracted
from the still widely-used plateau method applied to ratios of three-
and two-point functions at least up to source-sink separations of
$\sim 1.5$\,fm. The use of summed insertions, while generally an
important tool in suppressing excited-state effects on hadron
structure quantities, cannot reliably exclude a residual bias, in
particular when comparing with the results of fits which include
excited states explicitly. Moreover, for the first time, we apply the
full framework of covariant baryonic chiral perturbation theory
\cite{Becher:1999he,Kubis:2000zd,Fuchs:2003qc,Bauer:2012pv}
to the simultaneous determination of the form factors near $Q^2=0$ and
at the physical pion mass. From a careful study of all relevant
systematic effects, we are able to give a full error budget. Our final
results for various charge radii and the anomalous magnetic moments
$\kappa$ are listed in eq.\,(\ref{eq:bestresults}) below. We observe
agreement with experiment within the accuracy of our calculation,
including systematic errors. However, the overall uncertainty is too
large to have an impact on the proton radius puzzle.

This paper is organized as follows: In section~\ref{sec_setup}, we
describe our lattice setup, including details of the ensembles used
and observables measured, as well as our evaluation of statistical
errors. In section~\ref{sec_excite}, we discuss the analysis methods
we employed to study and suppress excited-state contributions. The
$Q^2$-dependence of the measured form factors, and the values of the
charge radii and magnetic moment determined from dipole fits on each
ensemble, are presented in section~\ref{sec_Qsqchiral}. In
section~\ref{sec_chpt}, we discuss in detail the chiral fits to the
form factors which we use to obtain our final results.
Section~\ref{conclusions} contains our conclusions and a brief
outlook.

A discussion of the impact of the use of Lorentz non-covariant
interpolating operators obtained from smearing the quark fields in the
spatial directions only on the Lorentz invariance of the results so
obtained is contained in appendix~\ref{app_smear}. For ease of
reference, we provide tables containing the full set of our results
for the form factors at all values of $Q^2$ on all ensembles in
appendix~\ref{wii}.

\section{Lattice setup \label{sec_setup}}

\subsection{Observables and correlators \label{subsec_observ}}

The matrix element of the electromagnetic current
\begin{equation}
  V^\mu_{\rm em}=\frac{2}{3} \bar{u}\gamma^\mu u -\frac{1}{3} \bar{d}
  \gamma^\mu d +\ldots
\end{equation}
between one-nucleon states can be expressed in terms of the Dirac and
Pauli form factors, $F_1$ and $F_2$. In Minkowski space notation the
form factor decomposition reads
\begin{equation}
\begin{multlined}
  \left\langle N(p^\prime,s^\prime)|V^\mu_{\rm em}(0)|N(p,s)
  \right\rangle=\\[0.2cm] 
  \bar{u}(p^\prime,s^\prime)\left[\gamma^\mu F_1(q^2)+
  i\frac{\sigma^{\mu\nu} q_\nu}{2m_{\rm N}}F_2(q^2)\right]u(p,s)\,,
\end{multlined}
\end{equation}
where $u(p,s)$ is a Dirac spinor with spin $s$ and momentum $p$,
$\gamma^\mu$ is a Dirac matrix,
$\sigma^{\mu\nu}=\frac{i}{2}\left[\gamma^\mu,\gamma^\nu\right]$, and
$m_{\rm N}$ denotes the nucleon mass.

The four-momentum transfer $q\equiv p^\prime-p$ is expressed in terms
of the energies and three-momenta of the initial and final states as
\begin{equation}
          q^2=-Q^2=(E_{\vec{p}^\prime}-E_{\vec{p}})^2-(\vec{p}^\prime
          - \vec{p})^2. 
\end{equation}
In this paper, we focus on the iso-vector form factors. By assuming
isospin symmetry, one can show via a simple
application of the Wigner-Eckart theorem applied in isospin space that
\begin{equation}
\begin{multlined}
  \left\langle {\rm p}(p^\prime,s^\prime)\right| \bar{u}\gamma^\mu u
  -\bar{d}\gamma^\mu d \left|{\rm p}(p,s) \right\rangle = \\[0.3cm]
  \left\langle {\rm p}(p^\prime,s^\prime) \right|V^\mu_{\rm em}
  \left| {\rm p}(p,s) \right\rangle -
  \left\langle {\rm n}(p^\prime,s^\prime) \right|V^\mu_{\rm em}
  \left| {\rm n}(p,s) \right\rangle,
\end{multlined}
\end{equation}
where $|{\rm p}\rangle$ and $|{\rm n}\rangle$ refer to one-proton and
one-neutron states, respectively. The expression on the left-hand side
is suitable for lattice QCD calculations, while the right-hand side
allows one to compare the results to experimental measurements.

The Dirac and Pauli form factors give rise to the helicity-preserving
and helicity-flipping contributions to the amplitude,
respectively. The electric and magnetic (Sachs) form factors $\GX{E}$
and $\GX{M}$ are obtained as linear combinations of $F_1$ and $F_2$,
\begin{eqnarray}
        \GX{E}(q^2)&=&F_1(q^2)+\frac{q^2}{4m_N^2}F_2(q^2)\,,\\
        \GX{M}(q^2)&=&F_1(q^2)+F_2(q^2)\,.
\end{eqnarray}
They can be determined from $ep$ scattering experiments by decomposing the
measured differential cross section through the Rosenbluth formula
\cite{Rosenbluth:1950yq}.
The form factors may be Taylor-expanded in the squared momentum
transfer $q^2$,
\begin{equation}
        \GX{E,M}(q^2)=\GX{E,M}(0)\left(1+\frac{1}{6}\langle r_{\rm E,M}^2
        \rangle q^2 +\mathcal{O}(q^4)\right),
\end{equation}
from which the charge radii of the nucleon may be determined:
\begin{equation}
        \langle r_{\rm E,M}^2 \rangle = \frac{6}{\GX{E,M}(q^2)}
        \frac{\partial \GX{E,M}(q^2)}{\partial q^2}\Bigg|_{q^2=0}.
\label{eq:rsqdef}
\end{equation}
Electric charge conservation implies $\GX{E}(0)=1$, while the magnetic
moment $\mu$ of the nucleon, in units of the nuclear magneton
$e/2m_{\rm N}$, is obtained from the magnetic form factor at vanishing
$q^2$, $\GX{M}(0)=\mu$.

\begin{figure}
\centering
        \includegraphics[width=0.42\linewidth]{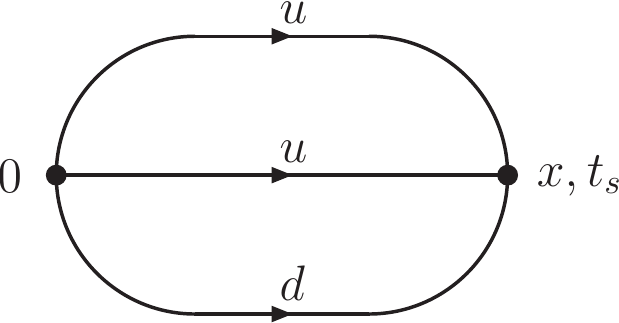}\qquad
        \includegraphics[width=0.42\linewidth]{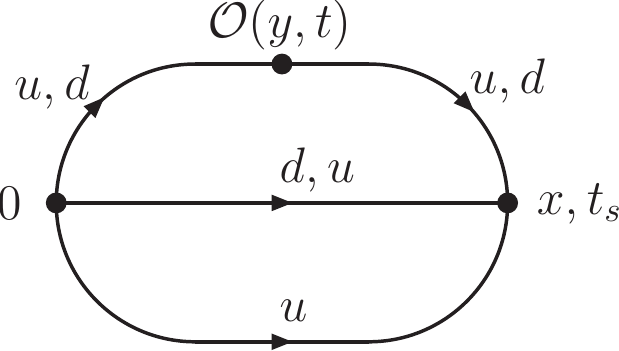}
        \caption{\label{fig:Feyn} \small Baryonic two-point and three-point
        functions (left and right panels respectively).}
\end{figure}

Lattice simulations allow for the determination of hadronic matrix elements by
computing Euclidean correlation functions of local composite
operators.\footnote{From here on we use Euclidean notation for the position-
  and momentum-space vectors, as well as for the Dirac matrices.} To
this end, one considers the nucleon two-point function
\begin{equation}
C_2(\vec{p},t) = \sum_{\vec{x}} \rme^{i\vec{p}\cdot\vec{x}}\,
\Gamma_{\beta\alpha}\,\langle \Psi^\alpha(\vec{x},t)\overline{\Psi}^\beta(0)\rangle, 
\end{equation}
where $\Psi^\alpha(\vec{x},t)$ denotes a standard interpolating operator for the
nucleon, and $\Gamma$ is a projection matrix in spinor space. The three-point
function of a generic (Euclidean) vector current $V_\mu$ is given by
\begin{equation}
C_{3,V_\mu}(\vec{q},t,t_s) = \sum_{\vec{x},\vec{y}}
\rme^{i\vec{q}\cdot\vec{y}}\Gamma_{\beta\alpha}\langle
\Psi^\alpha(\vec{x},t_s)V_\mu(\vec{y},t)\overline{\Psi}^\beta(0)\rangle\,,
\end{equation}
where $\vec{q}=\vec{p}^\prime-\vec{p}$. For the nucleon correlation functions
considered in this work, the projection matrix 
\begin{equation}
\Gamma=\frac{1}{2}(1+\gamma_0)(1+i\gamma_5\gamma_3)
\end{equation}
ensures the correct parity of the created states and gives the nucleon a
polarization in the $z$-direction, which is required to extract the magnetic
form factor. In the above expression for the three-point function the vector
current is inserted at Euclidean time~$t$, while the Euclidean time separation
between the initial and final nucleons is denoted by
$t_s$. Figure\,%
\ref{fig:Feyn}
shows the corresponding diagrams of the two- and three-point
functions. Note that for the iso-vector vector current considered in
this work, quark-disconnected diagrams cancel. Moreover, our
kinematics is chosen such that the final nucleon is always at rest,
i.e.
\begin{equation}
\vec{p}^\prime=0,\quad \vec{q}=-\vec{p}.
\end{equation}
The electric and magnetic form factors are easily determined from suitable
ratios of correlation functions. Here we follow ref.~%
\cite{Alexandrou:2008rp}
and use the ratio found to be most effective in isolating the desired matrix
element. For our chosen kinematics it reads
\begin{equation}
\begin{multlined}
        R_{V_\mu}(\vec{q},t,t_s)=\\
        \frac{C_{3,V_\mu}(\vec{q},t,t_s)}{C_2(\vec{0},t_s)}
        \sqrt{\frac{C_2(\vec{q},t_s-t)C_2(\vec{0},t)C_2(\vec{0},t_s)}
        {C_2(\vec{0},t_s-t)C_2(\vec{q},t)C_2(\vec{q},t_s)}} \,.
\label{eq:Ratio}
\end{multlined}
\end{equation}
From the asymptotic behaviour of $R_{V_\mu}(\vec{q},t,t_s)$ one can then
extract $\GX{E}$ and $\GX{M}$ for space-like momenta $Q^2\equiv-q^2>0$ via
\begin{equation}
        R_{V_0}(\vec{q},t,t_s)
        \stackrel{t,(t_s-t)\gg0}{\longrightarrow}\;
        \sqrt{\frac{m_N+E_{\vec{q}}}{2E_{\vec{q}}}}\GX{E}^{\rm bare}(Q^2), 
\label{eq:RatioV0}
\end{equation}
and
\begin{equation}
\begin{multlined}
        {\rm Re\,}R_{V_i}(\vec{q},t,t_s)
        \stackrel{t,(t_s-t)\gg0}{\longrightarrow}\;\\
        \epsilon_{ij3}q_j\sqrt{\frac{1}{2E_{\vec{q}}(E_{\vec{q}}+m_N)}}
        \GX{M}^{\rm bare}(Q^2),
\label{eq:RatioVi}
\end{multlined}
\end{equation}
where the $\epsilon$-symbol in the last equation denotes the
antisymmetric tensor with $\epsilon_{123}=+1$, and the superscripts
``bare'' remind us that, in general, the vector current requires
renormalization in the lattice-regularized theory.

\subsection{Simulation details \label{subsec_simul}}

Our calculations have been performed on a set of ensembles with
$N_f=2$ flavours of $\mathcal{O}(a)$-improved Wilson quarks and the
Wilson plaquette action. For the improvement coefficient $c_{\rm sw}$
we used the non-perturbative determination of ref.
\cite{impr:csw_nf2}.
The gauge configurations have been generated as part of the CLS
(Coordinated Lattice Simulations) initiative, using the
deflation-accelerated DD-HMC
\cite{Luscher:2005rx,Luscher:2007es}
and MP-HMC
\cite{Marinkovic:2010eg}
algorithms. Table~%
\ref{tab:ensembles}
provides details of the lattice ensembles used. 

\begin{table*}[t]
\begin{center}
\begin{ruledtabular}
\begin{tabular}{cccccccccccc}
        Run & $L/a$ & $\beta$ & $\kappa$ & $a m_\pi$ & $a m_N$ &  
        $m_\pi L$ & $N_\mathrm{cfg}$ & $N_\mathrm{meas}$ & $a$ 
        $[\mathrm{fm}]$ & $m_\pi$ $[\mathrm{MeV}]$ & $m_\pi / m_N$
         \\
        \hline
        $A3$ & 32 & 5.20 &  0.13580 & 0.1893(6) & 0.546(7) & 6.0 & 133  & 2128 & 0.079 &  $473$ & 0.346(5)\\
        $A4$ & 32 & 5.20 &  0.13590 & 0.1459(7) & 0.488(13) & 4.7 & 200  & 3200 & 0.079 &  $364$ & 0.299(7)\\
        $A5$ & 32 & 5.20 & 0.13594 & 0.1265(8) & 0.468(7) & 4.0 & 250   & 4000 & 0.079 &  $316$ & 0.270(5)\\
        $B6$ & 48 & 5.20 & 0.13597 & 0.1073(7) & 0.444(5) & 5.0 & 159   & 2544 & 0.079 &  $268$ & 0.242(3)\\
        \hline
        $E5$  & 32 & 5.30 & 0.13625 & 0.1458(3) & 0.441(4) & 4.7 & 1000 & 4000 & 0.063 &  $457$ & 0.330(3)\\
        $F6$  & 48 & 5.30 & 0.13635 & 0.1036(3) & 0.382(4) & 5.0 & 300    & 3600 & 0.063 &  $324$ & 0.271(3)\\
        $F7$  & 48 & 5.30 & 0.13638 & 0.0885(3) & 0.367(5) & 4.2 & 250    & 3000 & 0.063 &  $277$ & 0.241(4)\\
        $G8$ & 64 & 5.30 & 0.13642 & 0.0617(3) & 0.352(6) & 4.0 & 348   & 4176 & 0.063 &  $193$ & 0.175(3)\\
        \hline
        $N5$ & 48 & 5.50 & 0.13660 & 0.1086(2) & 0.329(2) & 5.2 & 477   & 1908 & 0.050 &  $429$ & 0.330(2)\\
        $N6$ & 48 & 5.50 & 0.13667 & 0.0838(2) & 0.297(3) & 4.0 & 946   & 3784 & 0.050 &  $331$ & 0.283(3)\\
        $O7$ & 64 & 5.50 & 0.13671 & 0.0660(1) & 0.271(4) & 4.4 & 490   & 1960 & 0.050 &  $261$ & 0.244(3)\\
\end{tabular}
\end{ruledtabular}
\end{center}
\caption{\label{tab:ensembles} Details of the lattice ensembles used
  in this study, showing the lattice extent, $L$, where $T=2L$; the
  values of the bare parameters $\beta$ and $\kappa$ in the lattice
  action; the pion and nucleon masses ($am_{\pi}$ and $am_N$); the
  number of measurements, $N_\textrm{meas}=N_\textrm{cfg}\times
  N_\textrm{src}$; the lattice spacing, $a$; the pion mass, $m_\pi$,
  in physical units, and the ratio $m_\pi/m_N$ of the pion and nucleon
  masses.}
\end{table*}

For the calculation of three-point correlation functions we employed
the point-split iso-vector current
\begin{equation}
\begin{multlined}
V_\mu^{\rm con}(x)=\frac{1}{2}
   \left(\psibar(x+\hat{\mu})(1+\gamma_\mu)U_\mu^\dagger(x)\tau^3\psi(x)
   \right.\\
        -\left.\psibar(x)(1-\gamma_\mu)U_\mu(x)\tau^3\psi(x+\hat{\mu})
        \right)\,,
\end{multlined}
\end{equation}
as well as the local vector current
\begin{equation}
V_\mu^{\rm loc}(x) = \psibar(x)\gamma_\mu\tau^3\psi(x).
\end{equation}
Here, $\psi$ denotes an isospin doublet of up and down quark fields,
and $\tau^3$ is the Pauli matrix acting in isospin space. While the
point-split current is conserved and satisfies the corresponding Ward
identity, the local vector current must be renormalized. The
expression for the renormalized current in the $\mathcal{O}(a)$
improved theory reads
\cite{Luscher:1996sc}
\begin{align}
V_\mu^\textrm{R}&=\zv(1+\bv a\mq)(V_\mu^{\rm loc}+a\cv\partial_\nu
T_{\mu\nu})\,, \label{eq:renorm_V} 
\end{align}
where $\mq$ denotes the bare subtracted quark mass, $\bv$ and $\cv$
are improvement coefficients, and $T^{\mu\nu}(x)= -\psibar(x)
\frac{1}{2} [\gamma_\mu,\gamma_\nu] \tau^3 \psi(x)$ is the tensor
density. We used the non-perturbative estimate for the renormalization
factor $\zv$ in the two-flavour theory of refs.
\cite{DellaMorte:2005rd,DellaMorte:2008xb}.
On the other hand, the conserved vector current, while not subject to
renormalization, requires $\mathcal{O}(a)$ improvement even at tree
level.  In this work we neither used the improved version of the
point-split vector current, nor did we compute matrix elements
containing the derivative of the tensor current. Therefore, our
results for form factors and charge radii are not fully
$\mathcal{O}(a)$ improved; hence, neglecting the $\bv$ term in
eq.\,(\ref{eq:renorm_V}) is consistent.

The interpolating field for the proton was chosen as
\begin{equation}
  \Psi^\alpha(x) = \epsilon_{abc} \left(u_a^T(x) C\gamma_5 d_b(x)
  \right) u^{\alpha}_c(x),
\end{equation}
with Gaussian-smeared quark fields
\cite{Gusken:1989ad}
\begin{equation}
  \widetilde\psi = \left(1+\kappa_{\rm G}\Delta\right)^N \psi\,,
\end{equation}
where the links in the three-dimensional covariant Laplacian $\Delta$
were APE-smeared
\cite{Albanese:1987ds}
in the spatial directions to further enhance the projection properties
onto the ground state and help reduce the gauge noise. Correlation
functions were constructed using identically smeared interpolating
fields at both source and sink to ensure that the two-point functions
are given by a sum of exponentials $e^{-E_n t}$ with positive
coefficients. The smearing parameter $\kappa_{\rm G}$ and the
iteration number $N$ were tuned so as to maximize the length of the
effective mass plateaux in a variety of channels. A widely used
measure for the spatial extent of a smeared source vector is the
``smearing radius'' $r_{\rm sm}$ (for a definition see e.g. eq.\,(2.6)
in
\cite{vonHippel:2013yfa}).
We note that our choice of $\kappa_{\rm G}$ and $N$ corresponds to
$r_{\rm{sm}}\approx 0.5$\,fm. As was first noted in
\cite{vonHippel:2013yfa},
the standard Gaussian smearing procedure becomes rapidly ineffective for
baryons as the lattice spacing is decreased. Alternatively one may employ
``free-form smearing''
\cite{vonHippel:2013yfa}
which, however, cannot be readily applied at the sink. Therefore, all results
presented in this paper have been obtained using standard Gaussian smearing at
both the source and sink. Note that we did not employ boosted Gaussian
smearing 
\cite{DellaMorte:2012xc}
either, because the boost is small for the nucleon, and the gain in terms of
an enhanced projection on the ground state is expected to be marginal.

Smearing the quark fields in the spatial directions only, while required in
order to keep the transfer matrix formalism intact, breaks the relativistic
covariance of the interpolating fields constructed from smeared quarks. This
issue has not been studied previously in any great detail in the
context of nucleon form factors. In appendix~%
\ref{app_smear}
we give a brief explanation why the relativistic invariance
of our results is not affected.

To compute the three-point function, we use the ``fixed-sink'' method,
which requires an additional inversion for each value of $t_s$, but
allows both the operator insertion and the momentum transfer to be
varied without additional inversions\,\cite{Martinelli:1988rr}. In
order to realize a range of values for the squared four-momentum
transfer~$Q^2$, we have computed the two- and three-point correlation
functions for several spatial momenta $\vec{q}\equiv \vec{n}2\pi/L$,
with $|\vec{n}|^2=0, 1, 2,\ldots, 6$.

The ratios $R_{V_\mu}$ of eq.~\eqref{eq:Ratio} contain a particular
combination of nucleon two-point functions, in order to isolate the
relevant matrix element. In our analysis the two-point functions which
enter $R_{V_\mu}$ were represented by single exponential fits. For
non-vanishing momenta $\vec{q}$, the nucleon energies were determined
from the nucleon mass using the continuum dispersion relation.
Compared to determining the nucleon energies directly from the
exponential fall-off of $C_2(\vec{q},t)$, we found that this procedure
resulted in smaller statistical errors, whilst producing compatible
results.

In order to express dimensionful quantities in physical units, we
determined the lattice spacing for all our ensembles using the mass of
the $\Omega$ baryon, as described in
\cite{Capitani:2011fg}.
More recently, the ALPHA collaboration has published accurate values
for the lattice spacing determined from the kaon decay constant, $f_K$
\cite{Fritzsch:2012wq,Lottini:2013rfa}.
While the central values differ slightly, both determinations are well
compatible within the quoted uncertainties. We have verified that
uncertainties in the scale-setting procedure have no significant influence on
the values of the charge radii in physical units.

\subsection{Statistics and error analysis\label{subsec_error}}

We computed two- and three-point correlation functions on all ensembles listed
in Table~%
\ref{tab:ensembles}.
In order to increase statistics, we used multiple sources spread
evenly across the lattice on each gauge configuration. The total
number of measurements for each ensemble is listed in Table~%
\ref{tab:ensembles}.
Statistical errors were estimated using a
bootstrap procedure with 10,000 bootstrap samples.

Simulations at the fine lattice spacings considered here are known to
be affected by the critical slowing-down of the smooth modes of the
gauge field and the freezing of the topological charge
\cite{Luscher:2011kk},
leading to potentially long autocorrelation times. Ignoring the long
tails in the autocorrelation function may lead to a significant
underestimation of statistical errors
\cite{Schaefer:2010hu}.
Since the correlation functions of the nucleon studied here are
intrinsically very noisy, however, one may expect that the
contributions from the tails have relatively little influence on the
overall statistical error.

We have investigated the impact of autocorrelations on our results by
performing a binning analysis prior to applying the bootstrap
procedure. To this end we focussed on the N6 ensemble, which is based
on a long Monte Carlo sequence, comprising 8040 molecular dynamics
units (MDUs) in total. Our findings indicate only a marginal increase
in the statistical error of the electric form factor and the nucleon
mass, which amounts to 2\% at most. We conclude that, for the purpose
of computing nucleon hadronic matrix elements and masses, our
ensembles are sufficiently decorrelated.

\section{Excited-state systematics \label{sec_excite}}

The standard ``plateau method'' for extracting $\GX{E}$ and $\GX{M}$ proceeds
by fitting the ratios defined in eq.~(\ref{eq:Ratio}) to a constant in the
region where they are approximately independent of $t$ and $t_s$, assuming
that their asymptotic behaviour has been reached. For the following discussion
it is useful to define an ``effective'' electric form factor, $\GX{E}^{\rm
  eff}(Q^2,t,t_s)$, by dividing out the kinematical factor in
eq.~(\ref{eq:RatioV0}), i.e.
\begin{equation}
  \GX{E}^{\rm eff}(Q^2,t,t_s) = \sqrt{\frac{2E_{\vec{q}}}{m_N+E_{\vec{q}}}}
  R_{V_0}(\vec{q},t,t_s).
\end{equation}
A similar relation is used to define $\GX{M}^{\rm eff}(Q^2,t,t_s)$. As $t,
(t_s-t)\to\infty$, the effective form factors will approach their asymptotic
values with exponentially small corrections,
\begin{equation}
  \GX{E,M}^{\rm eff}(Q^2,t,t_s) = \GX{E,M}(Q^2)
  +\mathcal{O}(e^{-\Delta t})+\mathcal{O}(e^{-\Delta^\prime(t_s-t)}),
\label{eq:FFsExc}
\end{equation}
where $\Delta$ and $\Delta^\prime$ denote the energy gaps between the ground
and first excited states for the initial- and final-state nucleons,
respectively. Here we omit the superscript ``bare'' on the form factors, since
we assume that the ratios $R_{V_\mu}$ have been appropriately renormalized.

\begin{figure*}
        \centering
        \includegraphics[width=.44\linewidth]{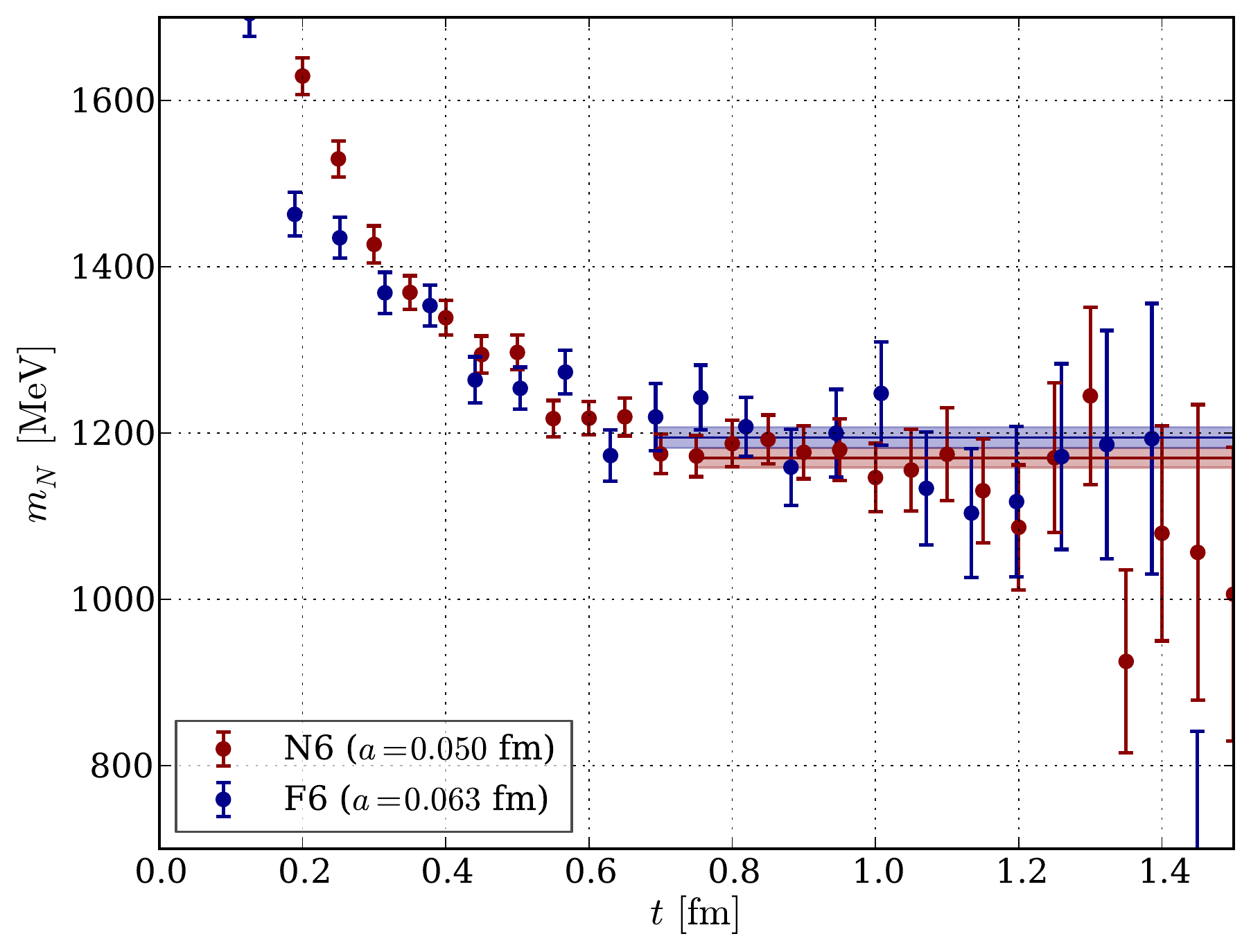}
        \hspace{0.8cm}
        \includegraphics[width=.44\linewidth]{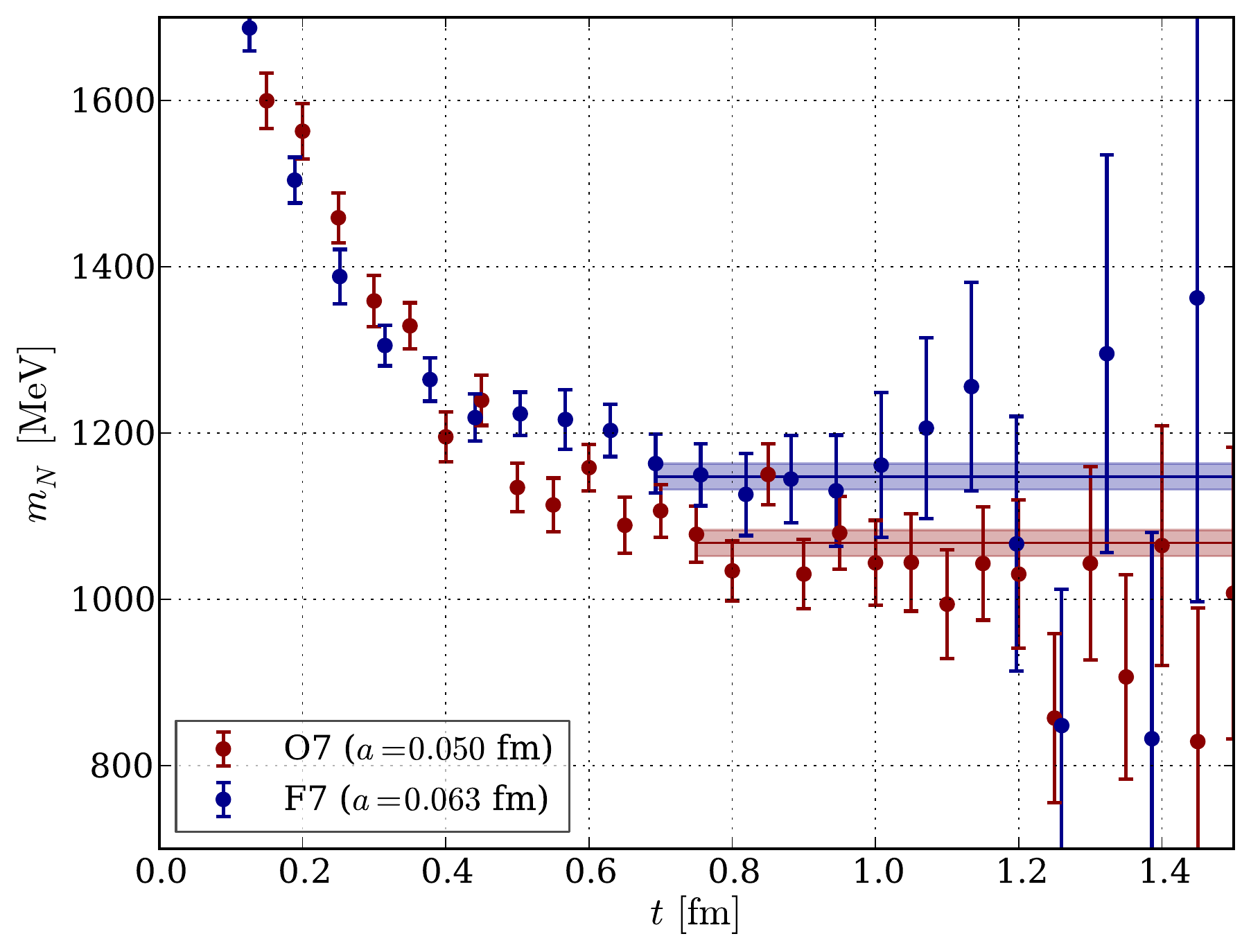}
        \caption{Effective masses in the nucleon channel computed at two
          different lattice spacings for
          $m_\pi\approx330$\,MeV (left panel) and $m_\pi\approx275$\,MeV
          (right panel). \label{fig:Meff_cmp}}
\end{figure*}

It is well known that nucleon correlation functions suffer from an
exponentially increasing noise-to-signal ratio
\cite{Parisi:1983ae,Lepage:1989hd}, which imposes a limit on the source-sink
separation $t_s$ which can be realized with reasonable numerical effort. In
typical calculations $t_s\approx1.1-1.2$\,fm, while separations as large as
1.4\,fm have been reported only in very few cases
\cite{Jager:2013kha,Green:2014xba}. Hence, in order to guarantee a reliable
determination of $\GX{E}$ and $\GX{M}$ using the plateau method, the
contributions from excited states in eq.~(\ref{eq:FFsExc}) must already be
sufficiently suppressed for $t, (t_s-t)\lesssim0.5$\,fm. Moreover, since the
gaps $\Delta$ and $\Delta^\prime$ are proportional to $m_\pi$ in the chiral
regime, one expects that this effect will become even more pronounced for the
more chiral ensembles.

In Fig.\,\ref{fig:Meff_cmp} we show effective mass plots for a nucleon at
rest, computed at two different values of the lattice spacing at nearly fixed
pion mass. One clearly sees that the ground state is isolated only for
separations larger than~0.5\,fm. Since the asymptotic behaviour must be
reached for both the initial and final-state nucleons, which may also carry
momentum, source-sink separations of the order of $1-1.5$\,fm seem rather
small. Therefore, one cannot rule out a systematic bias, unless source-sink
separations significantly larger than 1\,fm are realized.

\begin{figure*}
        \centering
        \includegraphics[width=.48\linewidth]{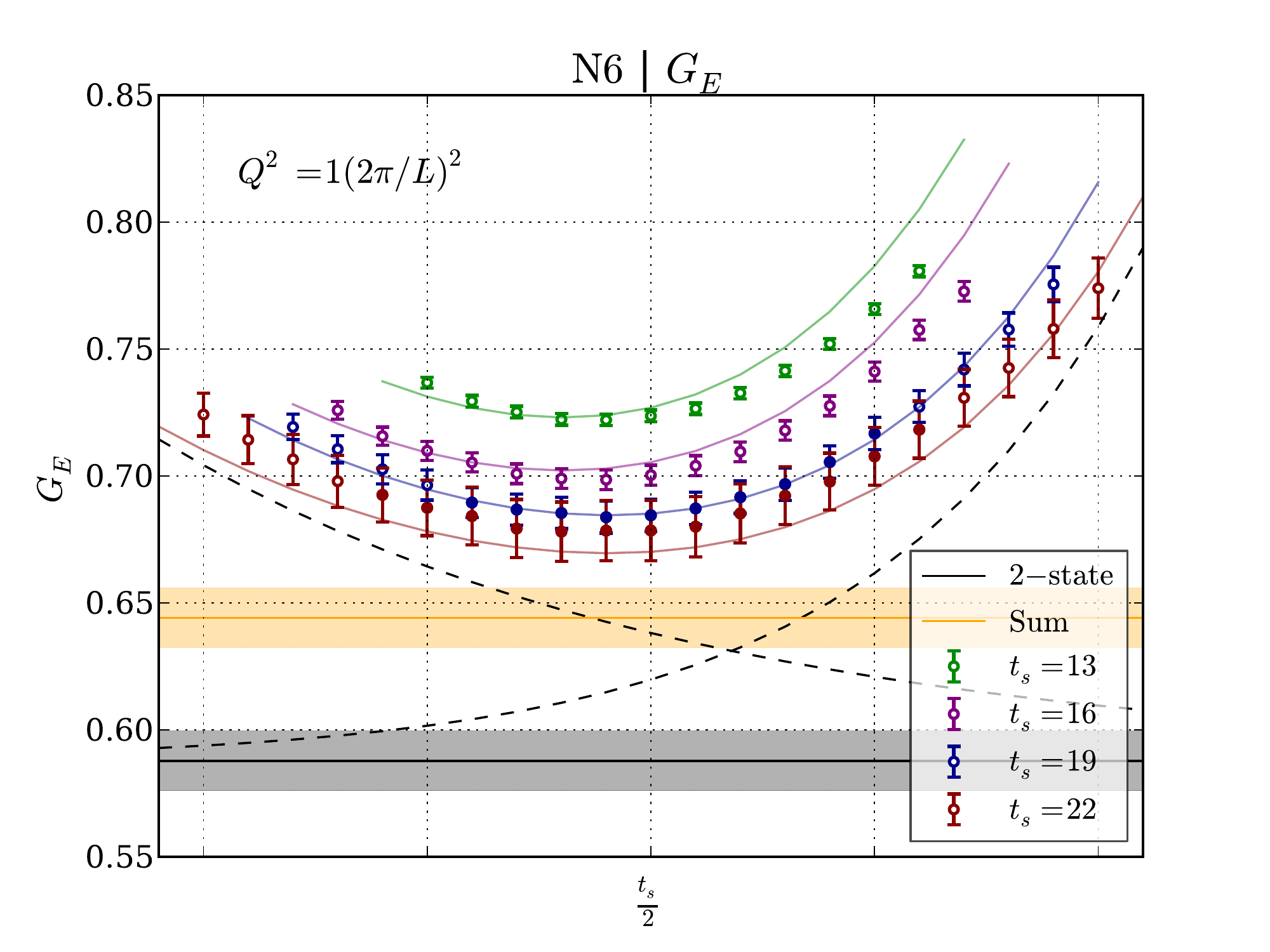}
        \hspace{0.4cm}
        \includegraphics[width=.48\linewidth]{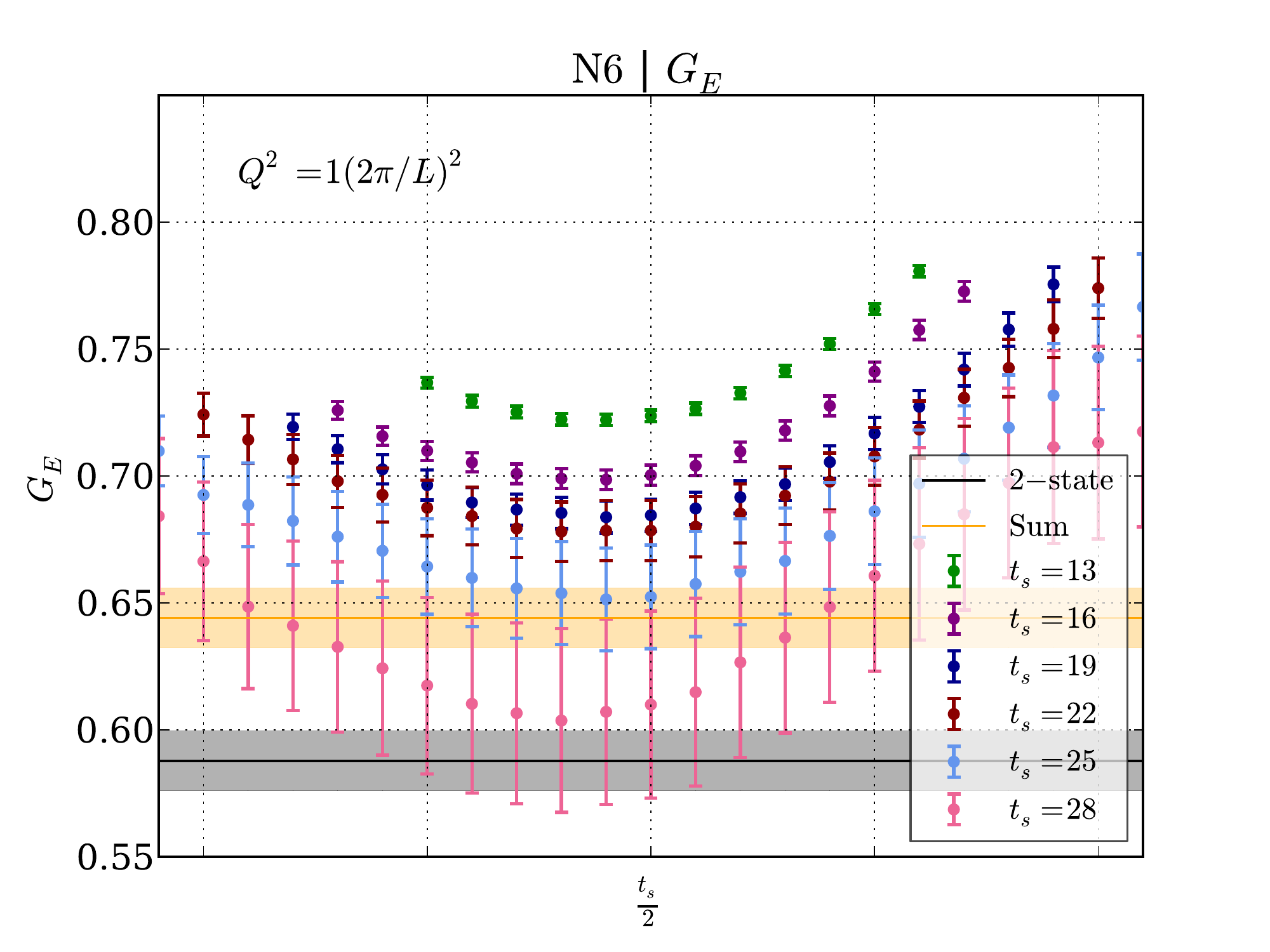}
        \caption{Data for $\GX{E}^{\rm eff}$ computed for several
          values of $t_s$ at the lowest non-zero momentum transfer on
          the N6 ensemble. The yellow band shows the result for
          $\GX{E}$ determined from the summation method. The solid
          curves are a representation of the data at individual values
          of $t_s$, as determined from a simultaneous two-state fit to
          the solid points in the left panel. The grey band denotes
          the corresponding asymptotic value. For the explanation of
          the dashed curves, see text. \label{fig:GE_cmp}}
\end{figure*}

In order to minimize or eliminate such a bias in our final results, we
employ three different methods:
\begin{itemize}
\item {\it Plateau fits:} For a fixed value of $t_s$ the quantities
  $\GX{E,M}^{\rm eff}(Q^2,t,t_s)$ are fitted to a constant over a
  small interval in~$t$. The default value of the source-sink
  separation is $t_s\approx 1.1$\,fm. For the high-statistics run on
  the N6-ensemble, we have also considered separations as large as
  $t_s=1.4$\,fm.
\item {\it Two-state fits:} In this case the leading contributions
  from excited states are included explicitly by using an {\it ansatz}
  of the form
  \begin{equation}
  \begin{multlined}
    \qquad \GX{E,M}^{\rm eff}(Q^2,t,t_s) = \GX{E,M}(Q^2) \\
    \qquad +c_{\rm E,M}^{(1)}(Q^2)\,e^{-\Delta t}
    +c_{\rm E,M}^{(2)}(Q^2)\,e^{-\Delta^\prime(t_s-t)},
  \end{multlined}
  \end{equation}
  with simultaneous fits in $t$ and $t_s-t$ performed to the data
  collected for several source-sink separations $t_s$. In order to
  stabilize the fits and reduce the number of fit parameters, we fix
  the gaps to $\Delta=m_\pi$ and $\Delta^\prime=2m_\pi$, assuming that
  the lowest-lying excitations are described by multi-particle states,
  consisting of a nucleon and at least one pion. In our chosen
  kinematics the nucleon at $t_s$ is at rest, such that the
  lowest-lying multi-particle state consists of one nucleon and two
  pions in an S-wave. By contrast, the initial state carries momentum
  and, in the absence of $\pi N$ interactions, therefore consists of a
  moving nucleon and a pion at rest, hence the choice
  $\Delta=m_\pi$. With these assumptions we may determine the form
  factors $\GX{E,M}$ as well as the coefficients $c_{\rm E,M}^{(1)}$
  and $c_{\rm E,M}^{(2)}$ as fit parameters for a given value of
  $Q^2$.
\item {\it Summed insertions (``summation method''):} Following
  refs.\,\cite{smear:Gaussian89,Maiani:1987by,Doi:2009sq,Bulava:2011yz}
  and our previous
  work\,\cite{Brandt:2011sj,Capitani:2012gj,Jager:2013kha} we define
  the quantities $S_{\rm E,M}(Q^2,t_s)$ by
  \begin{equation}
    S_{\rm E,M}(Q^2,t_s):=
    a\sum_{t=a}^{t_s-a} \GX{E,M}^{\rm eff}(Q^2,t,t_s),
  \end{equation}
  whose asymptotic behaviour is given by
  \begin{equation}
    \qquad S_{\rm E,M}(Q^2,t_s)\stackrel{t_s\gg0}{\longrightarrow}
    K_{\rm E,M}(Q^2) +t_s\,\GX{E,M}(Q^2) +\ldots,
  \label{eq:summedratio}
  \end{equation}
  where  $K_{\rm E,M}(Q^2)$ denote (in general divergent) constants,
  and the ellipses stand for exponentially suppressed corrections. The
  precise form of the latter depends on the details of the
  spectrum. If, for instance, $\Delta=m_\pi$ and
  $\Delta^\prime=2m_\pi$, the leading correction is of the order
  $\exp\{-\Delta t_s\}$, while for $\Delta=\Delta^\prime$ it is of the
  generic form 
  \begin{equation}
    (A_{\rm E,M}+B_{\rm E,M}t_s)\exp\{-\Delta t_s\},
  \end{equation}
  with coefficients $A_{\rm E,M}$ and $B_{\rm E,M}$. By computing
  $S_{\rm E,M}(t_s)$ for several sufficiently large values of $t_s$,
  form factors can be determined from the slope of a linear fit. Since
  $t_s$ is, by design, larger than either $t$ or $(t_s-t)$,
  excited-state contributions are parametrically reduced compared to
  the plateau method. The summation method has been successfully
  applied in our earlier calculation of the nucleon axial
  charge\,\cite{Capitani:2012gj} and also in recent studies of various
  nucleon matrix elements\,\cite{Green:2012ud,Green:2014xba}.
\end{itemize}

As a common feature among lattice calculations, we note that nucleon
electromagnetic form factors are typically overestimated at a given
value of $Q^2$ relative to the phenomenological representation of the
experimental data\,\cite{Kelly:2004hm}, even when the calculation is
performed for small pion masses. The three methods which we employ to
determine $\GX{E,M}(Q^2)$ are compared in Fig.\,\ref{fig:GE_cmp}. Our
data computed for different source-sink separations $t_s$ show a
systematic downward trend as $t_s$ is increased from~0.65
to~1.1\,fm. This reinforces our concern that $t_s\approx1.1$\,fm is
insufficient to rule out a systematic bias when the plateau method is
applied. In comparison, the slope determined from the summed ratio in
eq.\,(\ref{eq:summedratio}) yields a result for $\GX{E}(Q^2)$ which
lies sufficiently below $\GX{E}^{\rm eff}(Q^2,t,t_s)$ for all
$t,~t_s$.

The result from the two-state fit is even smaller: The asymptotic
value of $\GX{E}^{\rm eff}(Q^2,t,t_s)$ for $t, (t_s-t)\to\infty$ is
represented by the grey band, while the dashed curves correspond to
$\GX{E}+c_{\rm E}^{(1)}\,e^{-m_\pi{t}}$ and $\GX{E}+c_{\rm
  E}^{(2)}\,e^{-2m_\pi(t_s-t)}$, where $\GX{E}$, $c_{\rm E}^{(1)}$ and
$c_{\rm E}^{(2)}$ are determined from the fit. At face value, the
sizeable gap between the result from the two-state fit and the data
for $\GX{E}^{\rm eff}$ suggest that the latter are far from the
asymptotic behaviour when $t_s\leq1.1$\,fm. In particular, there is no
overlap between the grey band and any of the data points from which it
is determined. The dashed lines in the plot suggest that the two-state
fit constrains $\GX{E}$ merely from the curvature in~$t$ at a given
$t_s$ and from the trend in the source-sink separation as the latter
is increased. To investigate this further we have added two more
values of $t_s$ to the N6 ensemble, corresponding to separations
of~1.3 and 1.4\,fm, respectively. The additional data for $\GX{E}^{\rm
  eff}$ are shown in the right panel of Fig.\,\ref{fig:GE_cmp}. In
spite of the large statistical error, it is clear that $\GX{E}^{\rm
  eff}$ approaches the asymptotic value extracted from the two-state
fit. We conclude that two-state fits applied to our data collected for
$t_s\leq1.1$\,fm should not simply be discarded, even though the fit
corresponding to the grey band in the left panel of
Fig.\,\ref{fig:GE_cmp} does not appear very convincing. We will thus
include such results in our subsequent analysis, but interpret them
with the necessary amount of caution. Consequently, our preferred
method for determining form factors remains the summation method.

\section{$Q^2$-dependence and chiral behaviour\label{sec_Qsqchiral}}

In this section we discuss the dependence of form factors on the squared
momentum transfer, $Q^2$, and their behaviour as the pion mass is tuned
towards its physical value. Here we focus on the more qualitative features and
defer a detailed discussion of chiral extrapolations based on baryonic ChPT to
section\,\ref{sec_chpt}. 

For the remainder of this paper, we concentrate on results obtained using the
conserved (point-split) vector current, noting that the local vector current
yields fully consistent results, provided that it is properly renormalized.
Moreover, the ratio of matrix elements
computed using the local and conserved currents provides an estimate of the
renormalization factor $\zv$, which we find to be in agreement with other
work~\cite{DellaMorte:2005rd}.

{
\renewcommand\arraystretch{1.2}

\begin{table}[h!]
\begin{center}
\begin{ruledtabular}
\begin{tabular}{cccccccc}
 Run & method & $\langle r_{\rm E}^2 \rangle$ $[\textrm{fm}^2]$  &
 $\langle r_{\rm M}^2 \rangle$ $[\textrm{fm}^2]$ & $\mu$ \\ 
 \hline
 \hline
A3 & plat &  0.310(14) & 0.355(29) & 4.22(17)\\ 
 & sum &  0.335(27) & 0.391(56) & 4.65(34)\\ 
 & two-state &  0.339(18) & 0.425(55) & 4.48(29)\\ 
 \hline
A4 & plat &  0.362(23) & 0.324(52) & 3.65(29)\\ 
 & sum &  0.462(55) & 0.190(63) & 3.15(38)\\ 
 & two-state &  0.453(35) & 0.216(71) & 3.42(43)\\ 
 \hline
A5 & plat &  0.413(26) & 0.395(59) & 3.69(31)\\ 
 & sum &  0.504(57) & 0.333(98) & 3.72(51)\\ 
 & two-state &  0.543(56) & 0.53(21) & 4.3(1.0)\\ 
 \hline
B6 & plat &  0.427(22) & 0.442(32) & 3.89(17)\\ 
 & sum &  0.581(89) & 0.42(10) & 4.11(49)\\ 
 & two-state &  0.585(56) & 0.69(18) & 4.90(62)\\ 
 \hline\hline
E5 & plat &  0.304(14) & 0.318(44) & 4.03(30)\\ 
 & sum &  0.336(22) & 0.373(69) & 4.21(45)\\ 
& two-state &  0.385(19) & 0.367(69) & 4.16(48)\\ 
 \hline
F6 & plat &  0.407(17) & 0.387(27) & 3.67(18)\\ 
 & sum &  0.451(23) & 0.502(49) & 4.15(25)\\ 
& two-state &  0.505(21) & 0.366(68) & 3.32(33)\\ 
 \hline
F7 & plat &  0.421(25) & 0.385(45) & 3.72(25)\\ 
 & sum &  0.431(29) & 0.446(62) & 4.17(31)\\ 
 & two-state &  0.518(35) & 0.54(24) & 3.57(80)\\ 
 \hline
G8 & plat &  0.463(25) & 0.505(58) & 4.08(30)\\ 
 & sum &  0.502(76) & 0.45(13) & 4.58(73)\\ 
 & two-state &  0.739(76) & 0.85(39) & 5.8(1.6)\\ 
 \hline\hline
N5 & plat &  0.333(12) & 0.314(24) & 3.85(16)\\ 
 & sum &  0.383(29) & 0.361(60) & 3.93(35)\\ 
 & two-state &  0.381(12) & 0.399(34) & 4.18(21)\\ 
 \hline
N6 & plat &  0.391(16) & 0.412(33) & 3.91(21)\\ 
 & sum &  0.448(19) & 0.336(33) & 3.88(21)\\ 
 & two-state &  0.571(22) & 0.535(91) & 4.11(43)\\ 
 \hline
O7 & plat &  0.396(19) & 0.399(37) & 3.45(19)\\ 
 & sum &  0.460(23) & 0.436(42) & 3.66(21)\\ 
 & two-state &  0.672(38) & 0.99(20) & 4.98(69)\\ 
\end{tabular}
\end{ruledtabular}
\end{center}
\caption{Electric and magnetic charge radii and magnetic moment
  as determined from the three methods on each of our ensembles.
  \label{tab:radii}} 
\end{table}

}

A full set of results for $\GX{E}(Q^2)$ and $\GX{M}(Q^2)$ obtained
from all three methods (i.e. plateau fits, summation method and
two-state fits) is presented in
Tables\,\ref{tab:Q2_A3}-\ref{tab:Q2_O7} in appendix\,\ref{wii}. In
order to describe the dependence on $Q^2$ we follow the standard
procedure of fitting the results for $\GX{E,M}$ using a dipole
{\it ansatz} motivated by vector meson dominance, i.e.
\begin{equation}
\begin{multlined}
  \GX{E}(Q^2) = \left(1+\frac{Q^2}{M_{\rm E}^2}\right)^{-2}, \\
  \GX{M}(Q^2) = \GX{M}(0)\cdot \left(1+\frac{Q^2}{M_{\rm M}^2}\right)^{-2}\,.
\end{multlined}
\end{equation}
Using the definition in eq.\,(\ref{eq:rsqdef}), the charge radii are then
obtained from
\begin{equation}
  \left\langle r_{\rm E,M}^2\right\rangle = \frac{12}{M_{\rm E,M}^2}.
\end{equation}
The ratio $\GX{M}(Q^2)/\GX{E}(Q^2)$ is related to the nucleon's magnetic
moment $\mu$ via
\begin{equation}
  \mu\equiv 1+\kappa = \left. \frac{\GX{M}(Q^2)}{\GX{E}(Q^2)}\right|_{Q^2=0}, 
\end{equation}
where $\kappa$ denotes the anomalous magnetic moment. The reciprocal ratio
$\GX{E}(Q^2)/\GX{M}(Q^2)$ is an interesting quantity regarding the discrepancy
between experimental determinations based on Rosenbluth separation and the
recoil polarization technique.

\begin{figure*}
        \centering
        \includegraphics[width=.48\linewidth]{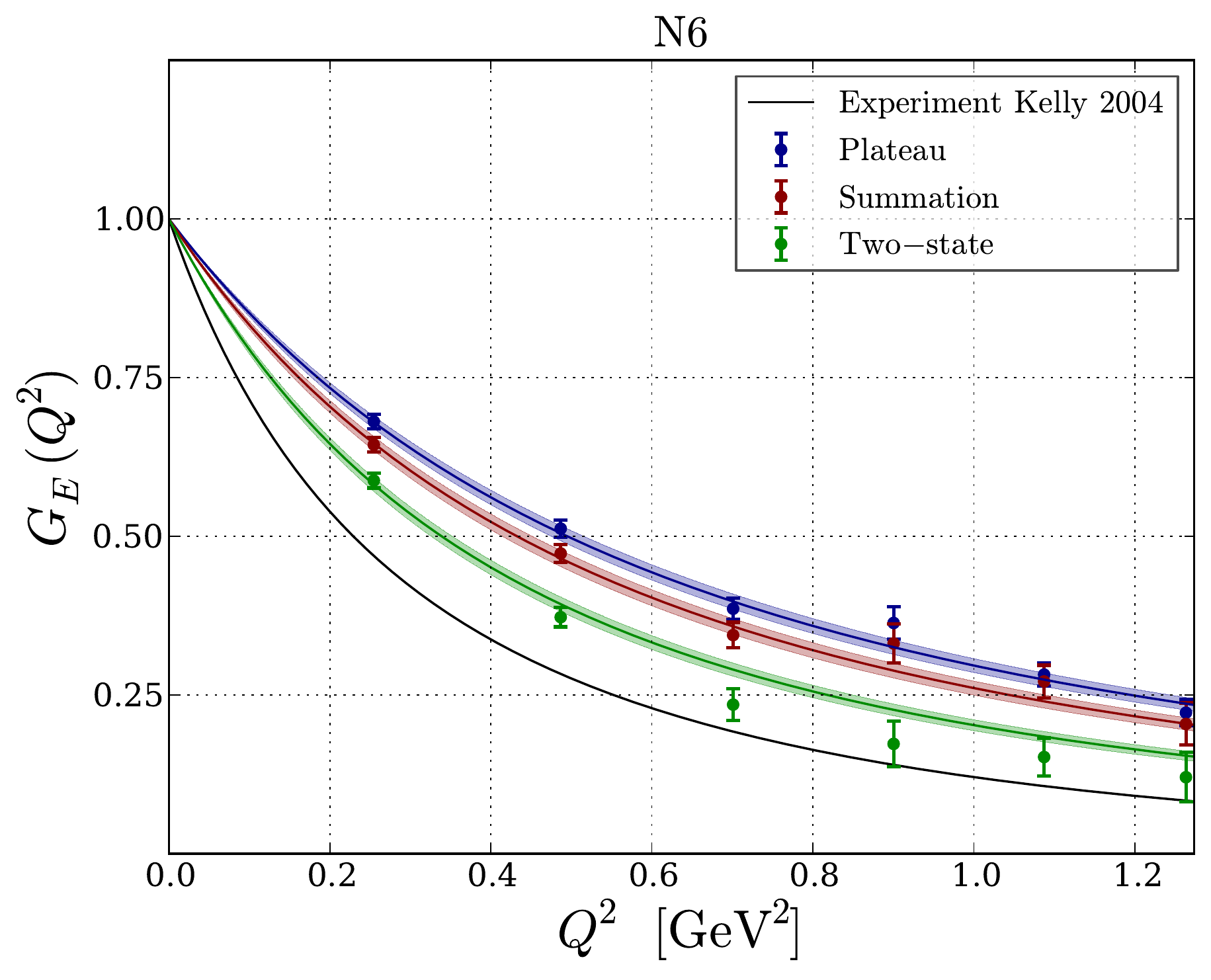}
        \includegraphics[width=.48\linewidth]{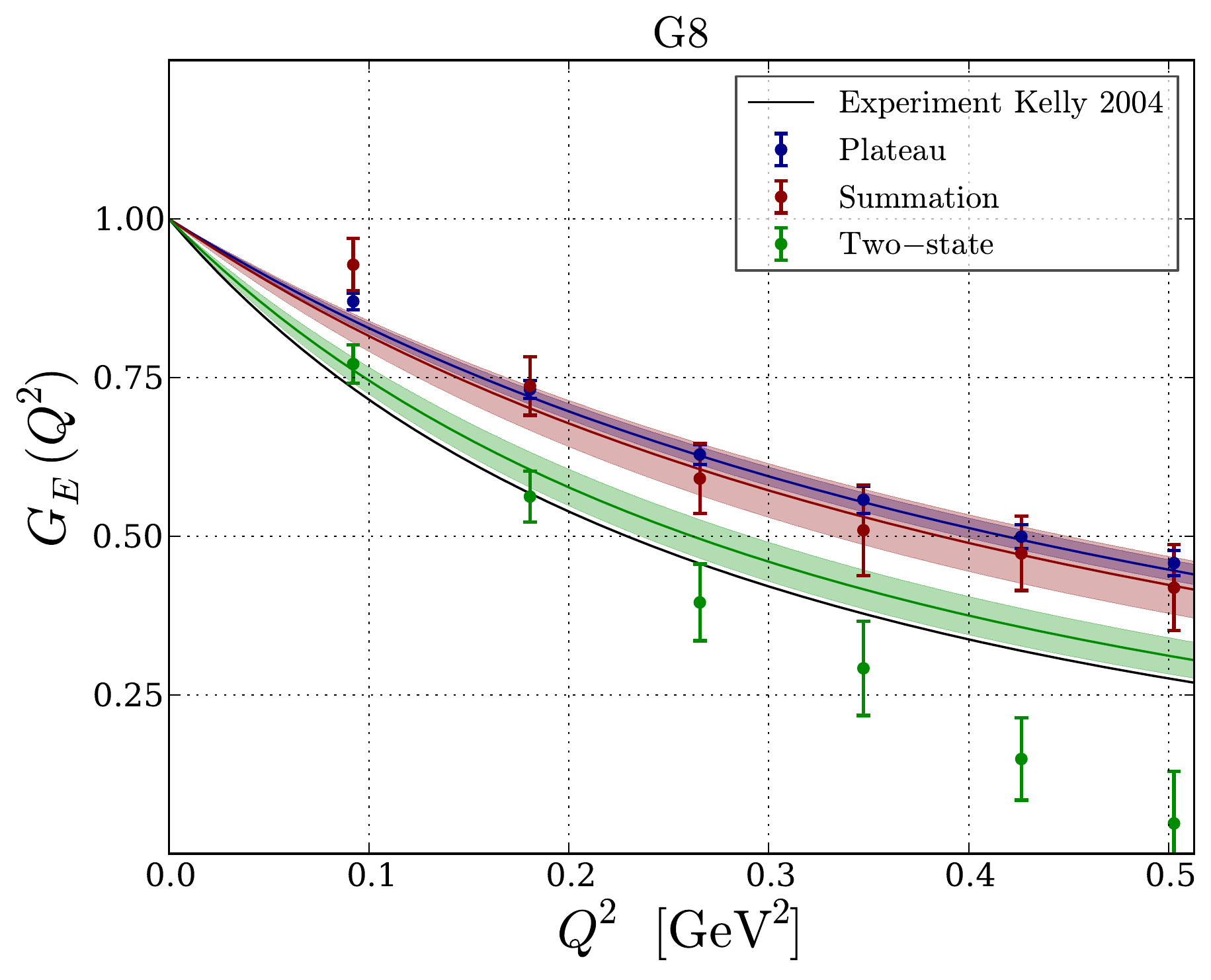}
        \caption{Dipole fits of the $Q^2$ dependence of $\GX{E}$, as
          determined using the plateau method (shown in blue), summed
          insertions (red) and two-state fits (green). The left and right
          panels correspond to pion masses of 331 and 193~MeV,
          respectively. The black line denotes Kelly's parameterization of
          experimental data.
\label{fig:GE_dip}}
\end{figure*}

\begin{figure*}
        \centering
        \includegraphics[width=.48\linewidth]{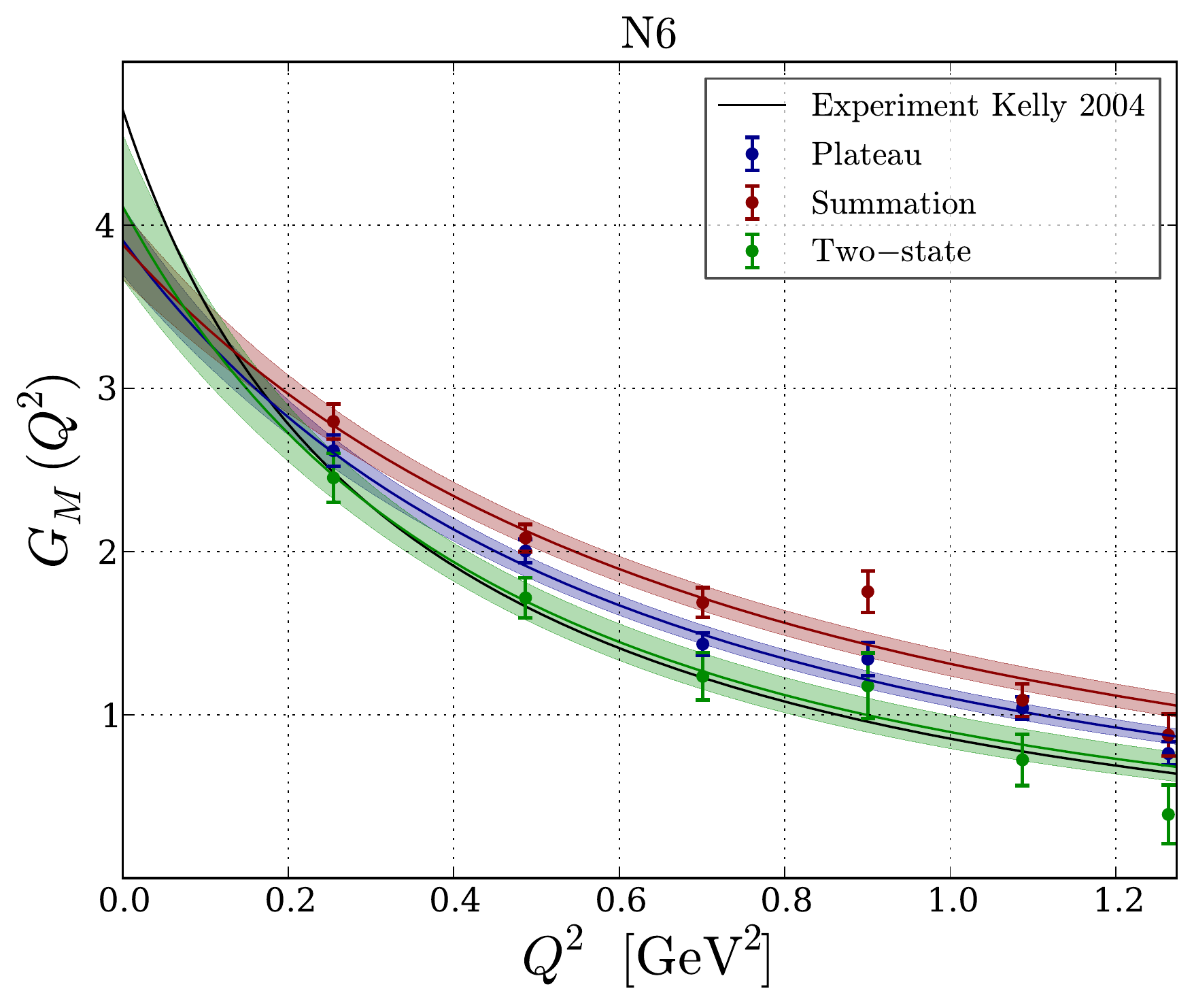}
        \includegraphics[width=.48\linewidth]{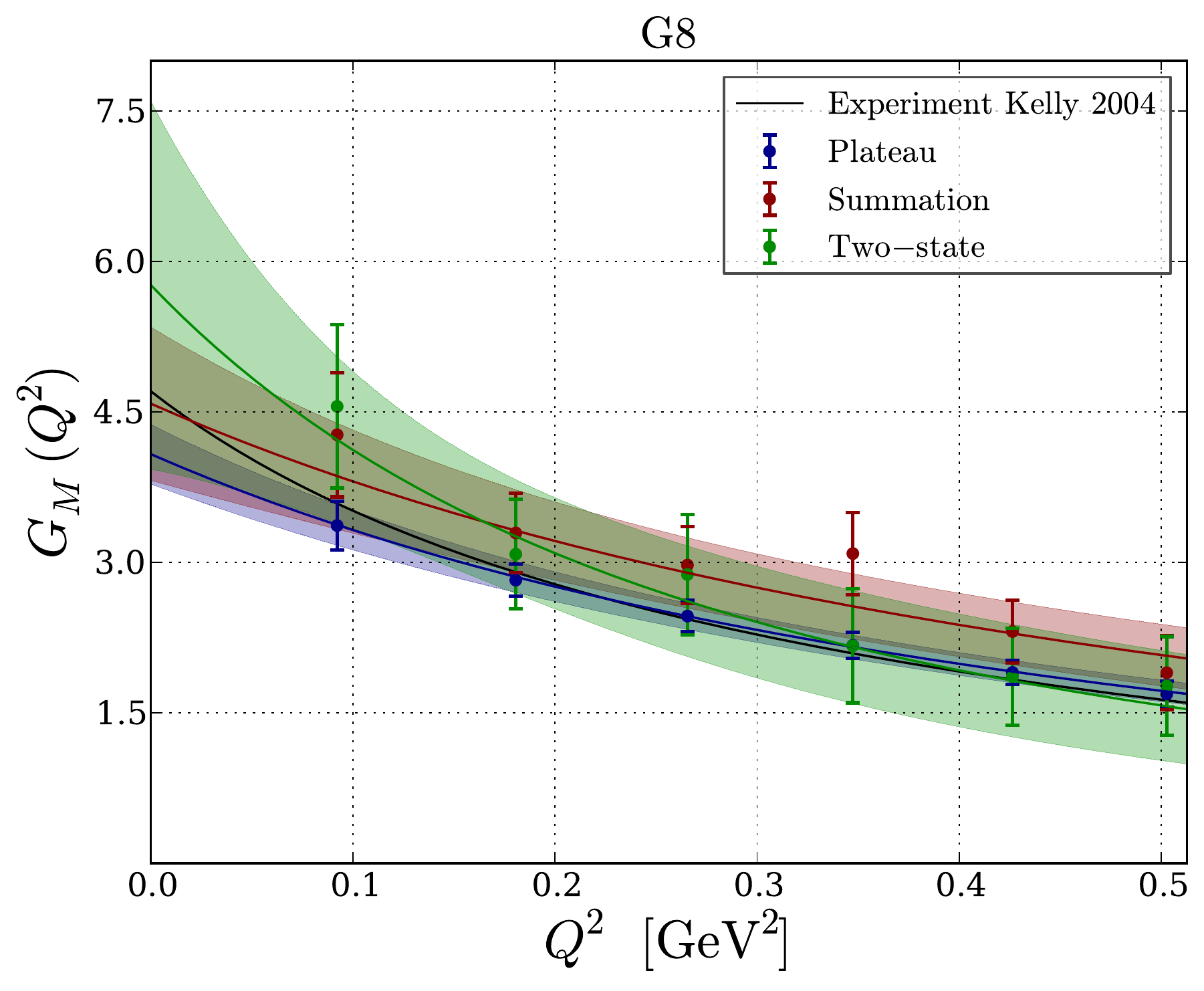}
        \caption{Dipole fits of the $Q^2$ dependence of $\GX{M}$. The meaning
          of the symbols is identical to Fig.\,\ref{fig:GE_dip}.
\label{fig:GM_dip}}
\end{figure*}

Examples of dipole fits to the form factor data obtained for two different
pion masses are shown in Figs.\,\ref{fig:GE_dip} and\,\ref{fig:GM_dip}, where
they are compared to Kelly's phenomenological
parameterization\,\cite{Kelly:2004hm} of experimental data.%
\footnote{We employ Kelly's parameterization as a benchmark, since 
the differences between Kelly and the more recent parameterization by
Arrington and Sick
\cite{Arrington:2006hm},
as well as the dispersive analysis by Lorenz {\it et al.}
\cite{Lorenz:2012tm}
are too small to be resolved at the level of statistical precision
provided by our data.}
In comparing experimental and lattice results, one must bear in mind
that the latter have been obtained at unphysical values of the pion
mass.

Clearly, the slope of the electric form factor near $Q^2=0$ varies
depending on the method which is used to determine $\GX{E}$ from the
ratio of correlators. One observes that two-state fits produce by far
the steepest drop-off, while the standard plateau method yields the
flattest behaviour in $Q^2$. This translates into a corresponding
hierarchy for estimates of the electric charge radius, which are
tabulated in Table~\ref{tab:radii}. As the pion mass is lowered
towards its physical value, one also finds that the spread in the
results for $\langle r_{\rm E}^2\rangle$ becomes more pronounced among
the three methods. This is consistent with the assertion that the
issue of unsuppressed excited state contributions becomes increasingly
important near the physical pion mass.

For $\GX{M}$ the systematic trend in the $Q^2$-dependence is not so
clearly visible as in the case of the electric form factor and charge
radius, which is partly due to the larger statistical
errors. Qualitatively, one observes that our lattice data for
$\GX{M}(Q^2)$ show better overall agreement with the representation of
the experimental data, regardless of the method which the former have
been obtained with.

\begin{figure*}
        \centering
        \includegraphics[width=.44\linewidth]{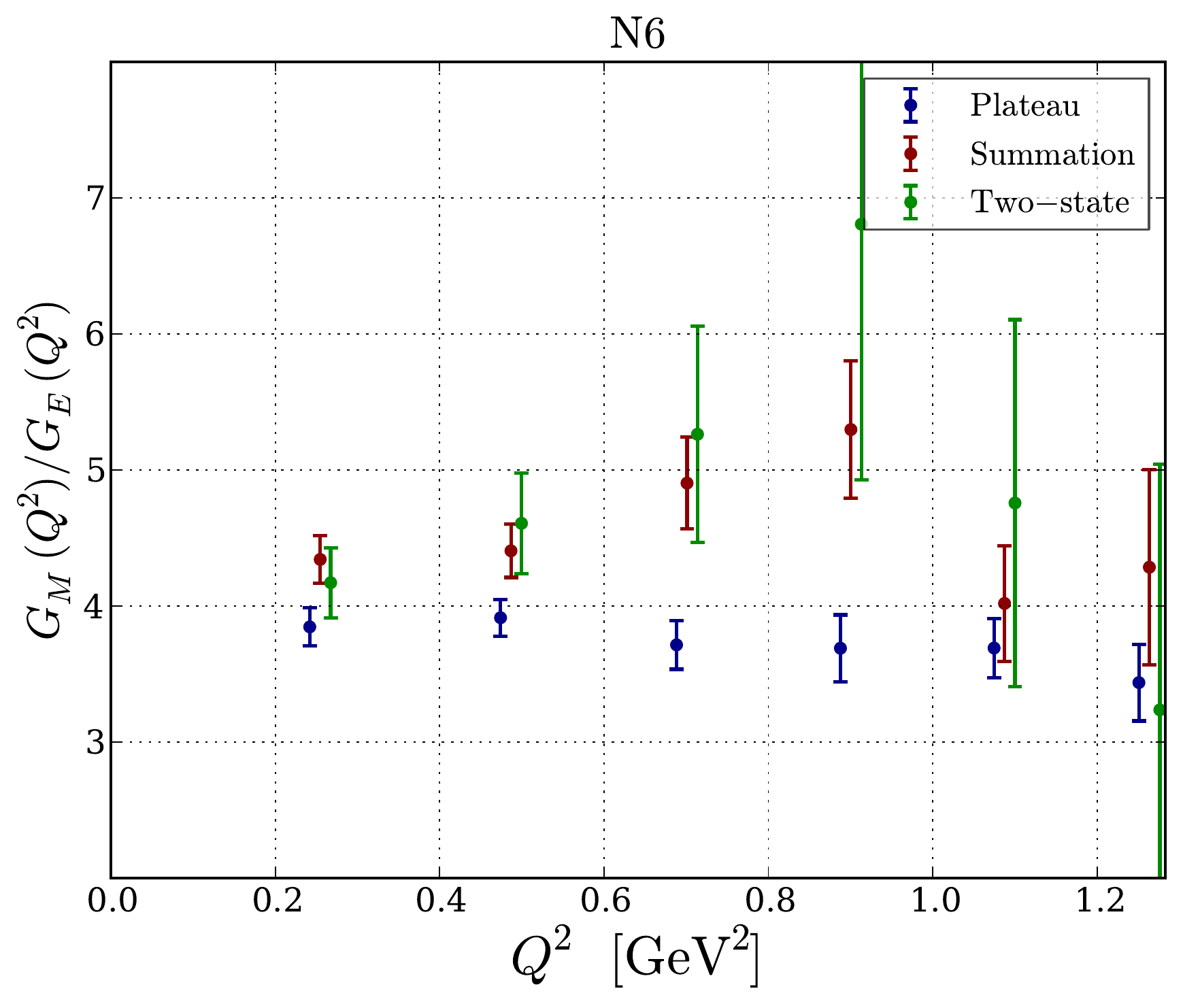}
        \hspace{0.8cm}
        \includegraphics[width=.44\linewidth]{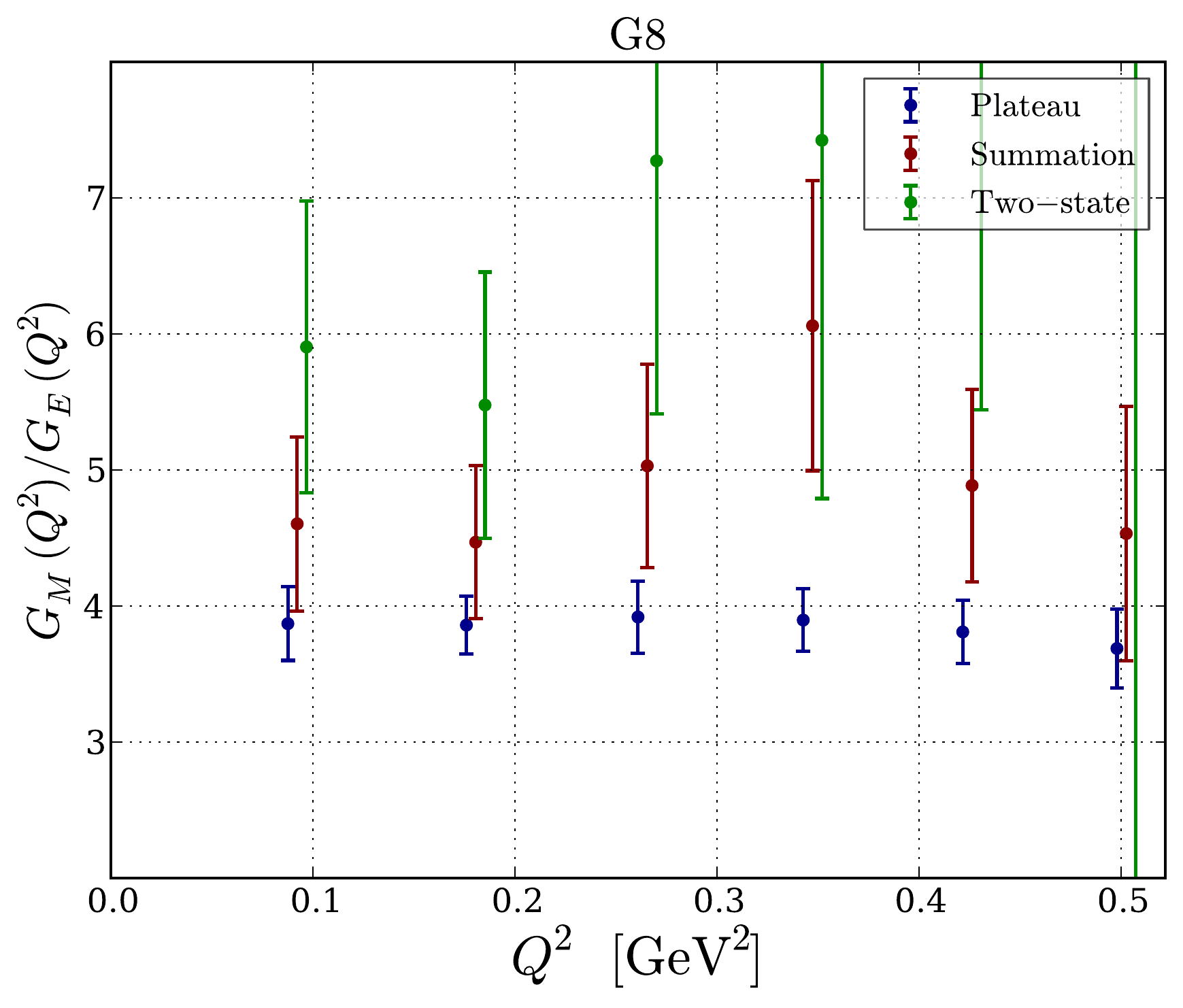}
        \caption{The $Q^2$-dependence of $\GX{M}/\GX{E}$.
\label{fig:mu}}
\end{figure*}

Dipole fits to the data for $\GX{M}(Q^2)$ extracted from two-state
fits show a slight -- albeit statistically insignificant -- tendency
for larger intercepts at vanishing $Q^2$, resulting in somewhat higher
estimates for the magnetic moment, $\mu$. Moreover, these fits reveal
that the electric and magnetic radii turn out to be rather similar
(see Table\,\ref{tab:radii}). The fact that $\GX{M}/\GX{E}$ shows no
statistically significant deviation from a constant within the
$Q^2$-range we are able to investigate is consistent with the
experimental data extracted using the traditional Rosenbluth
separation technique.

\begin{figure*}
        \centering
        \includegraphics[width=0.48\linewidth]{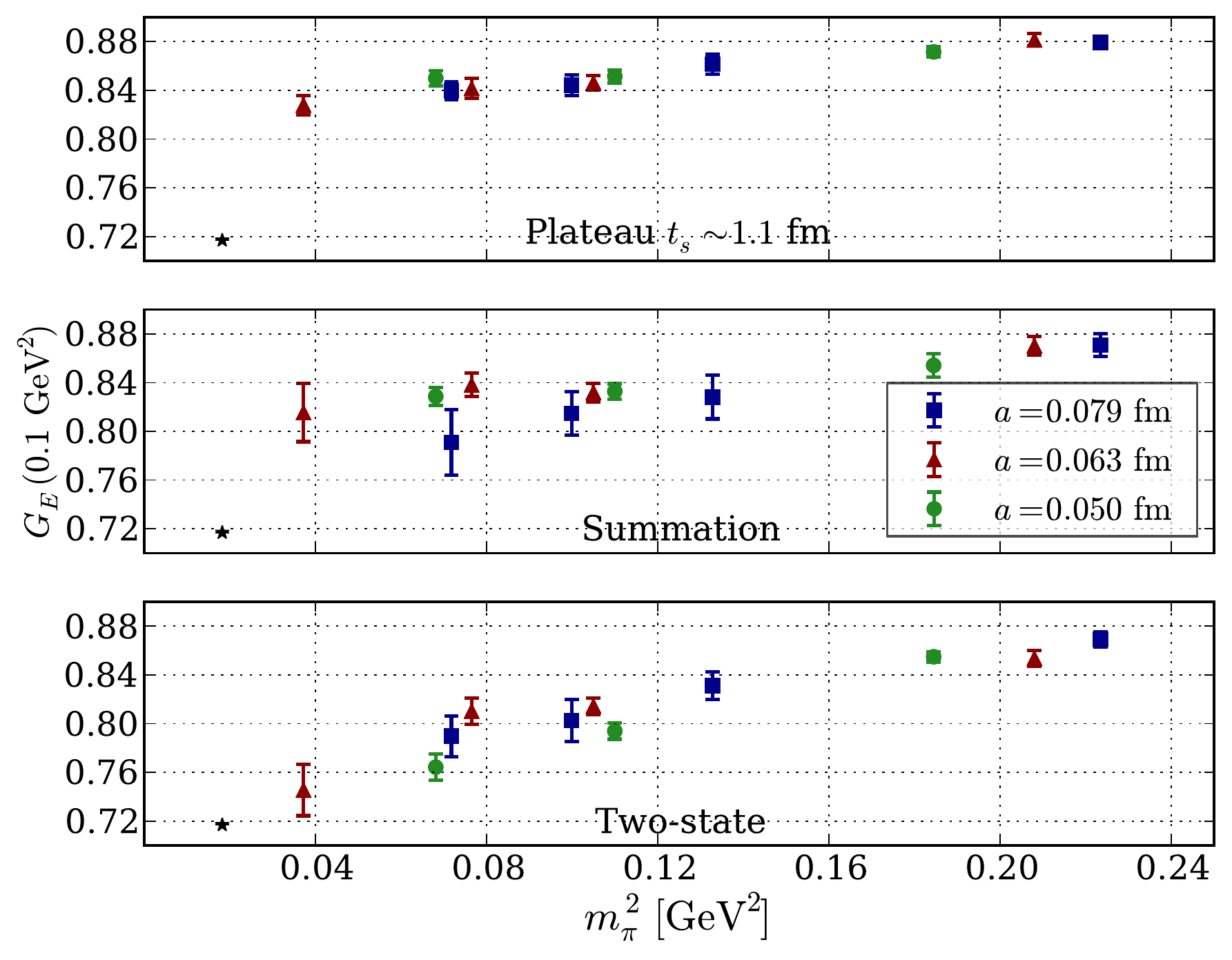}
        \hfill
        \includegraphics[width=0.48\linewidth]{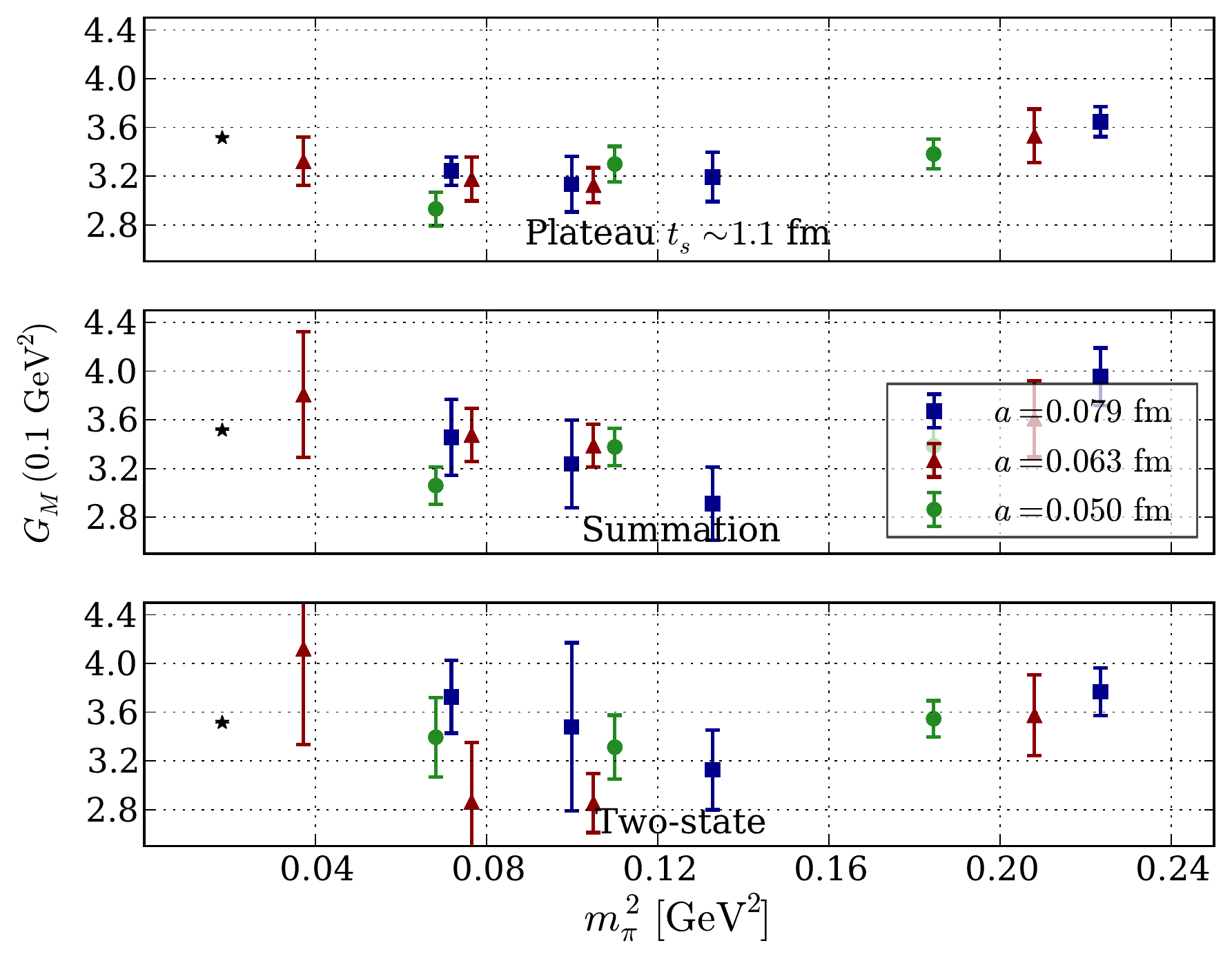}
        \caption{Comparison of the three methods for extracting the nucleon
          form factors $\GX{E}$ (left) and $\GX{M}$ (right)
          at a fixed value of $Q^2$. Shown are, from top to bottom,
          the results from the plateau method, summation method, and two-state
          fit, as a function of $m_\pi^2$. \label{fig:FFs_Qref}}
\end{figure*}

In order to further assess the effectiveness of the three methods employed to
extract the form factors, we have studied their chiral behaviour at a
reference value of the four-momentum transfer of
$Q_{\rm{ref}}^2=0.1\,\Gev^2$. This is very close to the smallest non-zero
value of $Q^2$ on the G8 ensemble, i.e. at our smallest pion mass. Using the
dipole fit parameters, we have obtained $\GX{E}(Q_{\rm{ref}}^2)$ and
$\GX{M}(Q_{\rm{ref}}^2)$ for all ensembles in our set. Similarly, we employed
the phenomenological parameterization to produce the corresponding estimates
from experiment at $Q_{\rm{ref}}^2$ and at the physical pion mass. The results
are shown in Fig.\,\ref{fig:FFs_Qref}. The plateau method clearly
overestimates $\GX{E}$, as there is no observable tendency for the data at
different pion masses to approach the experimental result. The summation
performs slightly better but does not improve the situation
substantially. Only the data based on two-state fits show a trend which brings
them into agreement with experiment at the physical pion mass. For the
magnetic form factor the situation is more favourable: the plateau method only
slightly underestimates $\GX{M}$ relative to experiment, while the chiral
trend in the data extracted using the summation method or two-state fits
agrees well with Kelly's parameterization.

We conclude that the summation method cannot fully reconcile lattice data for
$\GX{E}$ with its phenomenological value. Such an agreement can only be
reached if one is willing to trust two-state fits.

\section{Chiral fits\label{sec_chpt}} 

Our task is now to make contact between lattice data for form factors
obtained for a range of pion masses and lattice spacings and the
quantities which describe key properties of the nucleon, namely the
charge radii and the magnetic moments. This link is provided by chiral
effective field theory. The approach which has so far been most widely
applied to perform chiral extrapolations of lattice results for these
quantities is based on Heavy Baryon Chiral Perturbation Theory
(HBChPT)\,\cite{Jenkins:1990jv}, supplemented by the inclusion of the
$\Delta$-resonance\,\cite{Bernard:1998gv,Gockeler:2003ay}.

Here we employ an alternative formalism, i.e. the manifestly
Lorentz-invariant version of baryonic
ChPT\,\cite{Becher:1999he,Kubis:2000zd,Fuchs:2003qc}, which has also
been extended to include the $\Delta$-resonance\,\cite{Ledwig:2011cx},
as well as vector mesons\,\cite{Schindler:2005ke,Bauer:2012pv}. Our
procedure resembles the strategy pursued in\,\cite{Bauer:2012pv} to
extract charge radii and magnetic moments from experimental data of
nucleon form factors. In particular, we focus on fitting the
dependence of the form factors $\GX{E}$ and $\GX{M}$ on the pion mass
and the squared momentum transfer $Q^2$ to the expressions of baryonic
effective field theory (EFT), including vector degrees of freedom. The
relevant EFT expressions for $\GX{E}$ and $\GX{M}$ have been
supplemented by terms which describe the dependence on the lattice
spacing~$a$. In this way we combine a simultaneous chiral and
continuum extrapolation with a fit to the $Q^2$-dependence of form
factors. In order to enable a comparison with the standard approach we
also perform fits to the pion mass dependence of charge radii and the
magnetic moment to several variants of HBChPT.

\begin{table*}[t]
\begin{center}
\begin{ruledtabular}
\begin{tabular}{c c l}
                 & Low-energy & \\
\rb{Interaction} & parameter  & \rb{Value} \\
\hline
${\cal{L}}^{(2)}$
     & $F$    & $F_\pi^{\rm exp}=92.2$\,MeV \\
     & $M_\pi^2$ & Lattice input  \\
\hline
${\cal{L}}_{\rho,\rm eff}$
     & $\mathring{m}_\rho$ & $m_\rho^{\rm exp}=775$\,MeV or lattice input\\
     & $g$ & $g=m_\rho/\sqrt{2}F_\pi=5.93$ from KSRF relation \\[0.5ex] 
${\cal{L}}_{\pi\rho}$
     & $d_x$ & Fit parameter \\[0.5ex]
${\cal{L}}_{\pi\rho N}$
     & $g$ & from KSRF relation \\ 
     & $G_\rho$ & Fit parameter \\
\hline
${\cal{L}}^{(1)}_{\pi N}$
     & $\mathring{m}_{\rm N}$ & $m_{\rm N}^{\rm exp}=938$\,MeV or lattice input\\
     & $\mathring{g}_{\rm A}$ & $g_{\rm A}^{\rm exp}=1.27$ \\[0.5ex]
${\cal{L}}^{(2)}_{\pi N}$
     & $c_6$ & Fit parameter \\
     & $c_7$ & $c_7$ does not contribute in the iso-vector case \\[0.5ex]
${\cal{L}}^{(3)}_{\pi N}$
     & $d_6$ & Fit parameter \\
     & $d_7$ & $d_7$ does not contribute in the iso-vector case \\
\hline
${\cal{L}}^{(1)}_{\pi\Delta}$
     & $\mathring{m}_\Delta$ & $m_\Delta^{\rm exp}=1210$\,MeV \\[0.5ex]
${\cal{L}}^{(1)}_{\pi N\Delta}$
     & $g_{\pi N\Delta}$ & $1.125$, from fit to $\Delta\to\pi N$ decay width \\
\end{tabular}
\end{ruledtabular}
\end{center}
\caption{Interaction terms of the chiral effective theory used for fitting
  lattice data for nucleon electromagnetic form factors. A circle on top a
  symbol indicates that the corresponding low-energy parameter is defined in
  the chiral limit. The third column describes how their values are
  determined. Further details are described in the text.}
\label{tab:VChPT}
\end{table*}

Specifically we consider the manifestly Lorentz-invariant effective
Lagrangian describing ${\pi}N$ interactions including vector mesons at
$\mathcal{O}(q^3)$ in the chiral expansion. A detailed discussion of
this effective Lagrangian can be found in\,\cite{Bauer:2012pv}.
Table\,\ref{tab:VChPT} gives an overview of the various interaction
terms, as well as the associated low-energy constants and hadron
masses. From the table one can read off which low-energy constants are
determined by our fitting procedure and which phenomenological
information is used to fix the values of the remaining ones. We note
that the interaction terms proportional to $c_7$ and $d_7$ do not
contribute in the iso-vector case considered here. We have also
dropped the contributions from the $\omega$-meson entirely, since they
were found to have only a negligible effect on the
results\,\cite{Bauer:2012pv}. Furthermore, in
ref.\,\cite{Djukanovic:2004mm} it was shown that the universal
$\rho$-meson coupling constant~$g$ can be fixed via the
Kawarabayashi-Suzuki-Riadzuddin-Fayyazuddin (KSRF)
relation\,\cite{Kawarabayashi:1966kd,Riazuddin:1966sw}, which follows
by requiring the self-consistency of an effective chiral theory
involving pion, nucleons and the $\rho$-meson.

The full expressions for the chiral expansions of the Dirac and Pauli
form factors to $\mathcal{O}(q^3)$ are given in appendix~D.2 of
ref.\,\cite{TBauer_dipl} but are too lengthy to be displayed here.
Starting from those formulae, we have formed the appropriate linear
combinations for the iso-vector form factors $\GX{E}$ and
$\GX{M}$. The resulting expressions were used to perform a
simultaneous fit to both $\GX{E}(Q^2)$ and $\GX{M}(Q^2)$ obtained for
a range of pion masses and momentum transfers, at all three values of
the lattice spacing.%
\footnote{To evaluate the loop integrals appearing in the expressions,
we make use of LoopTools~\cite{Hahn:1998yk,vanOldenborgh:1989wn}.}
Cutoff effects can be easily incorporated into this
framework by adding terms proportional to the lattice spacing~$a$ to
the form factors, i.e.
\begin{eqnarray}
 &&   \GX{E}(Q^2) = \GX{E}^{\rm EFT}(Q^2)+aQ^2\,\beta_{\rm E},
  \nonumber \\ 
 &&  \GX{M}(Q^2) = \GX{M}^{\rm EFT}(Q^2)+a\,\beta_{\rm M},
\label{eq:cutoff}
\end{eqnarray}
where $\GX{E,M}^{\rm EFT}$ denote the continuum EFT expressions for
the form factors, while the coefficients $\beta_{\rm E,M}$ are taken
as fit parameters. This {\it ansatz} takes account of the fact that
the matrix element corresponding to the electric form factor is
$\mathcal{O}(a)$-improved at vanishing momentum transfer.

Estimates for the charge radii $\langle r_{\rm E}^2\rangle$, $\langle
r_{\rm M}^2\rangle$ and the anomalous magnetic moment $\kappa$ are
obtained by inserting the fitted values of the low-energy parameters
of $d_x, G_\rho, d_6$ and $\tilde{c}_6\equiv c_6-G_\rho/2g$ into the
corresponding EFT expressions, i.e.
\begin{eqnarray}
 &&  \left\langle r_{\rm E,M}^2 \right\rangle = -\frac{6}{\GX{E,M}(0)}
    \left.\frac{d\GX{E,M}(Q^2)}{dQ^2}\right|_{Q^2=0}, \\
 && \kappa = \GX{M}(0)-1.
\end{eqnarray}
The relations between these quantities and the Dirac radius $\langle
r_1^2\rangle$, as well as the combination $\kappa\langle r_2^2\rangle$
reads
\begin{equation}
\begin{multlined}
   \left\langle r_1^2\right\rangle = \left\langle r_{\rm E}^2
   \right\rangle -\frac{6\kappa}{4m_{\rm N}^2}, \\
   \kappa \left\langle r_2^2 \right\rangle = (1+\kappa) \left\langle
   r_{\rm M}^2 \right\rangle - \left\langle r_{\rm E}^2
   \right\rangle +\frac{6\kappa}{4m_{\rm N}^2}
\label{eq:EM_vs_DiracPauli}
\end{multlined}
\end{equation}

We refer to a fit applied to form factor data obtained from the
summation method, over the entire range of $Q^2$, with a pion mass cut
of $m_\pi \leq m_\pi^{\rm cut}=330$\,MeV, and the masses of the
$\rho$-meson and nucleon fixed to their experimental values as our
standard procedure. Standard fits were performed with and without
terms parameterizing lattice artefacts (see
eq.\,(\ref{eq:cutoff})). For the fit including lattice artefacts one
finds $\chi^2_{\rm red}=1.21$, for 66 degrees of freedom. Results for
the quantities $\langle r_{\rm E}^2\rangle$, $\langle r_{\rm
  M}^2\rangle$, $\kappa$, $\langle r_{1}^2\rangle$ and the combination
$\kappa\langle r_{2}^2\rangle$ are listed in
Table\,\ref{tab:errorbudget}. One observes that fits with and without
${\cal{O}}(a)$ terms produce compatible results: differences at the
level of at most 1.5~standard deviations are seen in $\langle r_{\rm
  M}^2 \rangle$ and $\kappa$. This indicates that the influence of
lattice artefacts on the results is small.

\begin{figure*}
        \centering
        \includegraphics[width=.44\linewidth]{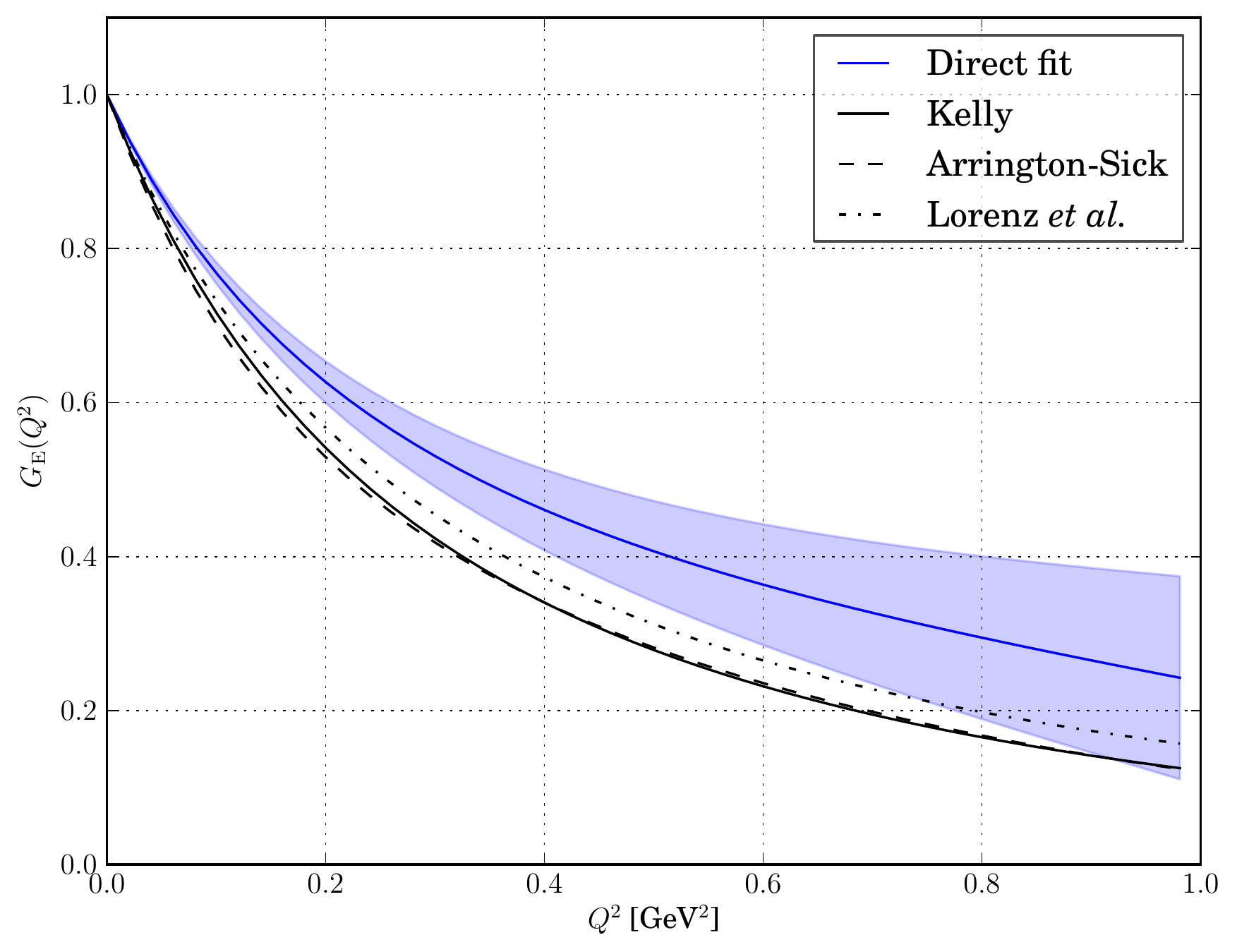}
        \hfill
        \includegraphics[width=.44\linewidth]{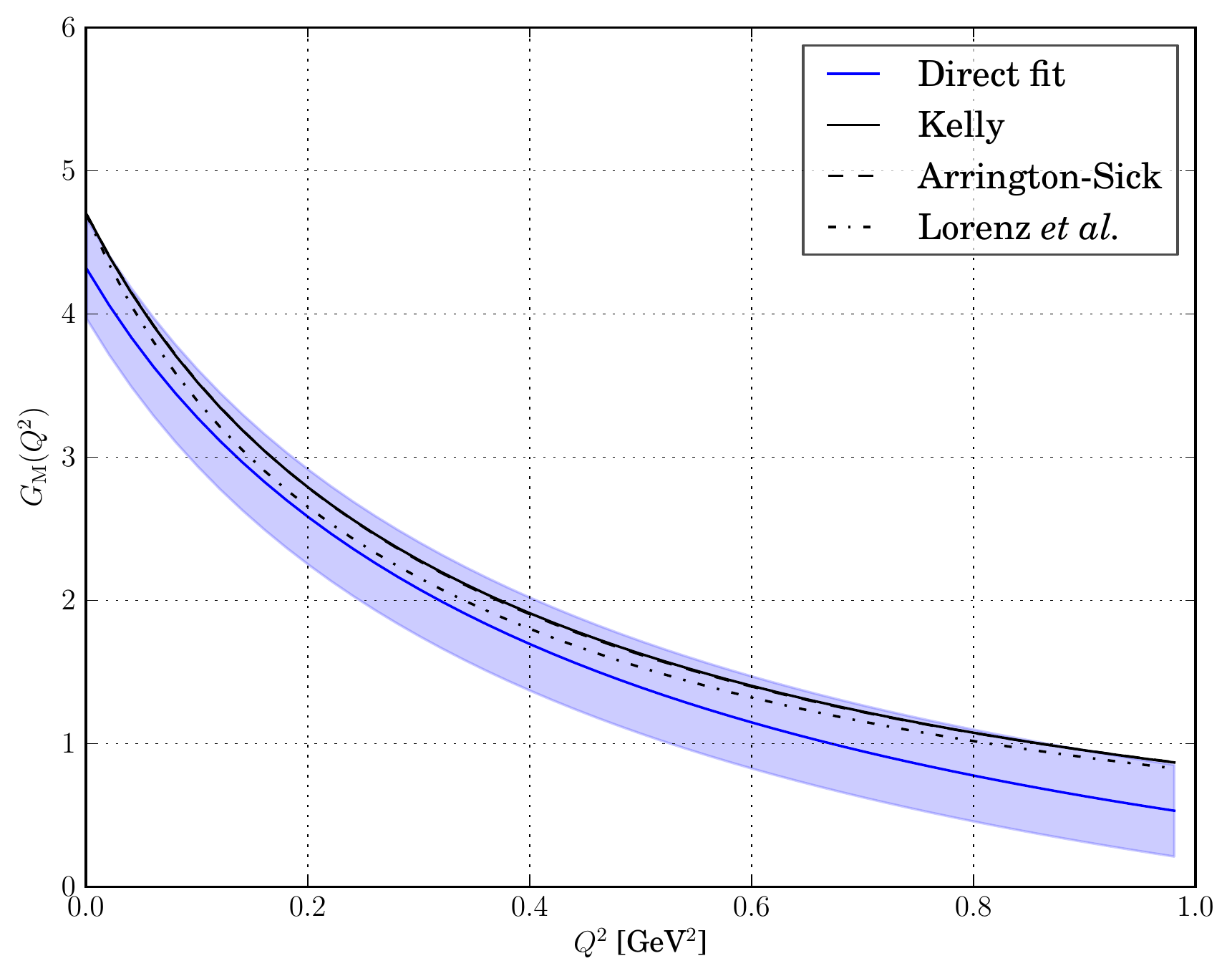}
        \caption{The $Q^2$ dependence of $\GX{E}$ and $\GX{M}$ at the
          physical pion mass and in the continuum limit, as determined
          from a simultaneous fit to lattice data (``direct'' fit)
          including lattice artefacts. The bands denote the
          statistical error. The solid, dashed and dashed-dotted
          curves are the phenomenological representations of
          experimental data of
          refs.\,\cite{Kelly:2004hm},\,\cite{Arrington:2006hm}
          and\,\cite{Lorenz:2012tm}, respectively.
\label{fig:Qsquared_phys}}
\end{figure*}

A first qualitative assessment can be made by plotting the
$Q^2$-dependence of $\GX{E}$ and $\GX{M}$ at the physical point
compared to various phenomenological parameterizations of experimental
data, as shown in Fig.\,\ref{fig:Qsquared_phys}. While the dependence
of $\GX{E}$ on the squared momentum transfer is somewhat flatter
compared to experiment, the behaviour of $\GX{M}$ is reproduced very
well.

\begin{table*}
\begin{center}
\begin{tabular}{cccccc}
\hline\hline
 $\langle r_{\rm E}^2\rangle [\fm^2]$ & $\langle r_{\rm
  M}^2\rangle [\fm^2]$ & $\kappa$ & $\langle r_{1}^2\rangle [\fm^2]$
 & $\kappa\langle r_2^2\rangle [\fm^2]$ & Fit \\
\hline
 0.722(34) & 0.720(53) & 3.33(35) & 0.501(41) & 2.61(9) & 
 standard, with ${\cal{O}}(a)$-terms \\ [1.0ex]
 0.748(12) & 0.636(8) & 3.93(11) & 0.487(14) & 2.65(9) & 
 standard, without ${\cal{O}}(a)$-terms \\ [1.0ex]
  ${-0.013}$ & ${-0.004}$ & ${-0.15}$ & ${-0.003}$ & ${-0.11}$ &
 variant 1 \\ [1.0ex] 
  ${+0.030}$ & ${-0.025}$ & ${+0.28}$ & ${+0.011}$ & ${+0.04}$ &
 variant 2 \\ [1.0ex] 
  ${-0.001}$ & ${-0.009}$ & ${-0.14}$ & ${+0.009}$ & ${-0.14}$ &
 variant 3 \\ [1.0ex] 
  ${+0.030}$ & ${+0.035}$ & ${-0.40}$ & ${+0.057}$ & ${-0.16}$ &
 variant 4 \\ [1.0ex] 
  ${+0.083}$ & ${+0.029}$ & ${-0.16}$ & ${+0.093}$ & ${-0.05}$ &
 variant 5 \\ 
\hline\hline
\end{tabular}
\end{center}
\caption{Results for charge radii and magnetic moments from direct
  fits to the form factors. Lines $3-7$ contain the differences
  between variants of the fitting procedure as labelled in
  Table\,\ref{tab:Fitvariants} and the results from the standard fit
  performed without ${\cal{O}}(a)$-terms. For instance, variant~5
  yields $\langle r_{\rm E}^2\rangle=0.831$.}
\label{tab:errorbudget}
\end{table*}

\begin{table*}
\begin{center}
\begin{tabular}{lclcc}
\hline\hline
Standard fit & & Variants & & Label \\
\hline
Impose pion mass cut of $m_\pi^{\rm cut}\leq 330$\,MeV
 & & $m_\pi^{\rm cut}\leq 300$\,MeV & & 1 \\
 & & No mass cut & & 2 \\ \hline
Fit entire available range in $Q^2$
 & & Impose cut of $Q^2<0.5\,\rm GeV^2$ & & 3 \\ \hline
Use experimental values for $m_{\rm N}, m_\rho$
 & & Use lattice input for $m_{\rm N}, m_\rho$ & & 4 \\ \hline 
Fit data obtained using the summation method
 & & Fit data extracted from two-state fits & & 5 \\ 
\hline\hline
\end{tabular}
\end{center}
\caption{The standard procedure for fitting nucleon form factors to
  the expressions from baryonic ChPT and the variants applied in order
  to estimate the systematic error.}
\label{tab:Fitvariants}
\end{table*}

In order to estimate the systematic error, we have considered a number
of variations in the fitting procedure, which are compiled and
labelled in Table\,\ref{tab:Fitvariants}. These include different pion
mass cuts, restrictions of the fitted range in $Q^2$ and the use of
the masses of the $\rho$ and nucleon determined by the lattice
calculation at the respective value of the pion mass. These variations
are indicative of higher-order terms in the chiral expansion and probe
the overall consistency of our particular EFT approach. We have also
estimated the residual systematic uncertainty due to excited states,
by repeating the entire procedure using the form factor data obtained
from two-state fits. Variations of the fitting procedure
(corresponding to the entries in lines $3-7$ in
Table\,\ref{tab:errorbudget}) were always applied neglecting terms
parameterizing lattice artefacts (i.e. for $\beta_{\rm E}=\beta_{\rm
  M}=0$), as this produced more stable fits, in particular when
imposing more aggressive cuts in the pion mass or $Q^2$
range. However, while the systematic error budget is estimated from
fits excluding lattice artefacts, we prefer to quote our main results
using fits in which ${\cal{O}}(a)$-terms have been accounted for.

We thus obtain as our final results:
\begin{eqnarray}
&&  \left\langle r_{\rm E}^2\right\rangle = 0.722\pm0.034\,({\rm stat})
               {}^{+0.030}_{-0.013}\,({\chi{\rm fit}})
               {}^{+0.083}_{-0.000}\,({{\rm exc}})\,\fm^2, \nonumber \\
&&  \left\langle r_{\rm M}^2\right\rangle = 0.720\pm0.053\,({\rm stat})
               {}^{+0.035}_{-0.025}\,({\chi{\rm fit}})
               {}^{+0.029}_{-0.000}\,({{\rm exc}})\, \fm^2, \nonumber \\
&&  \kappa = 3.33\pm0.35\,({\rm stat})
               {}^{+0.28}_{-0.40}\,({\chi{\rm fit}})
               {}^{+0.00}_{-0.16}\,({{\rm exc}}), \label{eq:bestresults} \\
&&  \left\langle r_{1}^2\right\rangle = 0.501\pm0.041\,({\rm stat})
               {}^{+0.057}_{-0.003}\,({\chi{\rm fit}})
               {}^{+0.093}_{-0.000}\,({{\rm exc}})\, \fm^2, \nonumber \\
&&  \kappa\left\langle r_2^2\right\rangle = 2.61\pm0.09\,({\rm stat})
               {}^{+0.04}_{-0.16}\,({\chi{\rm fit}})
               {}^{+0.00}_{-0.05}\,({{\rm exc}})\, \fm^2. \nonumber
\end{eqnarray}
Here, the systematic uncertainties estimated from fit variants $1-4$
have been combined into an overall chiral fitting error, while the
difference between employing the summation method and two-state fits
is quoted as a separate, residual systematic uncertainty arising from
excited states.

We did not consider fits to baryonic EFT including $\Delta$ degrees of
freedom when assessing our systematic errors, as such fits produced
unacceptably large values of $\chi^2_{\rm red}$ when the low-energy
parameter $g_{\pi N\Delta}$ was fixed to the phenomenological value
of~1.125. On the other hand, treating $g_{\pi N\Delta}$ as a fit
parameter resulted in an unphysically small value.

\begin{figure*}
        \centering
        \includegraphics[width=.46\linewidth]{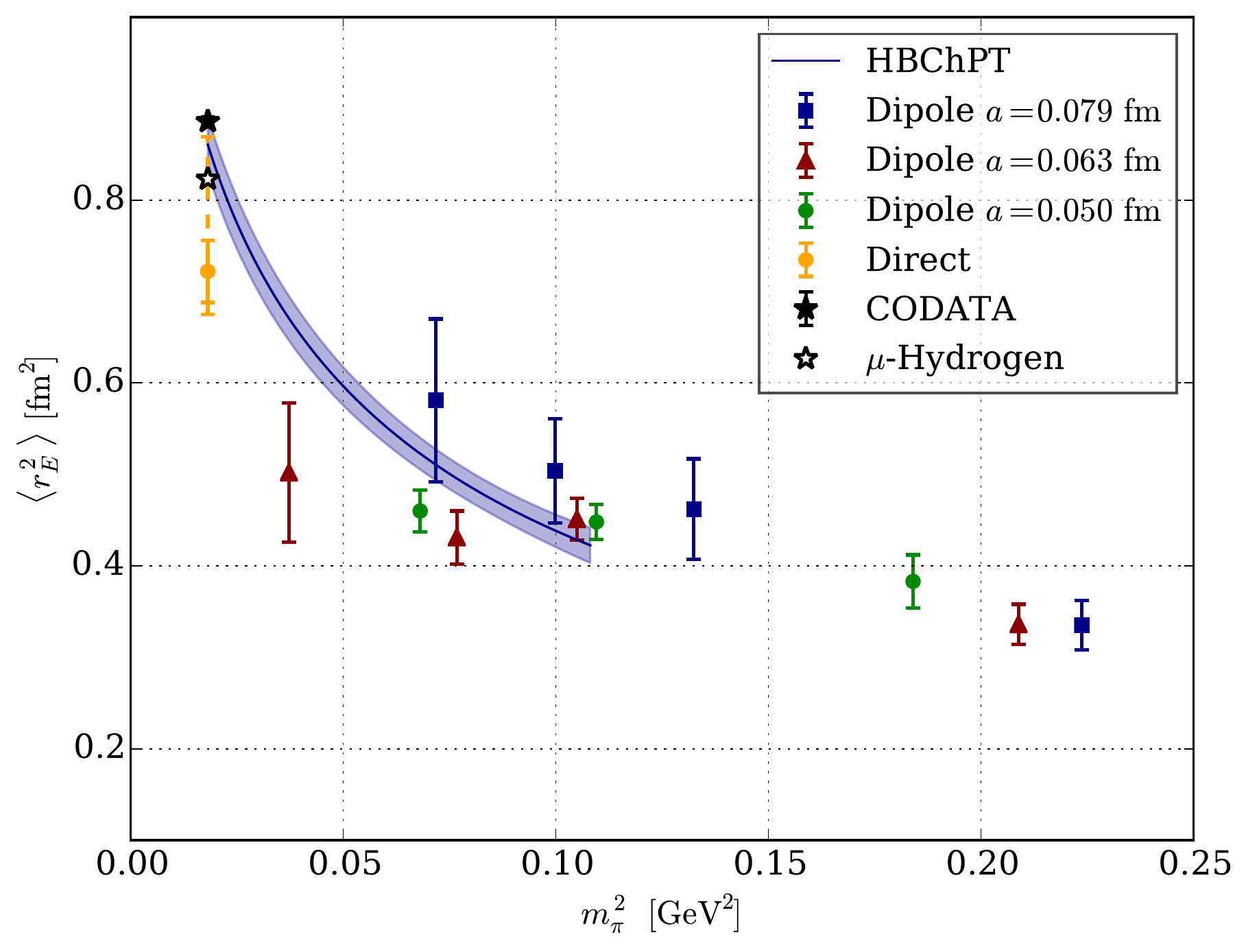}
        \hfill
        \includegraphics[width=.46\linewidth]{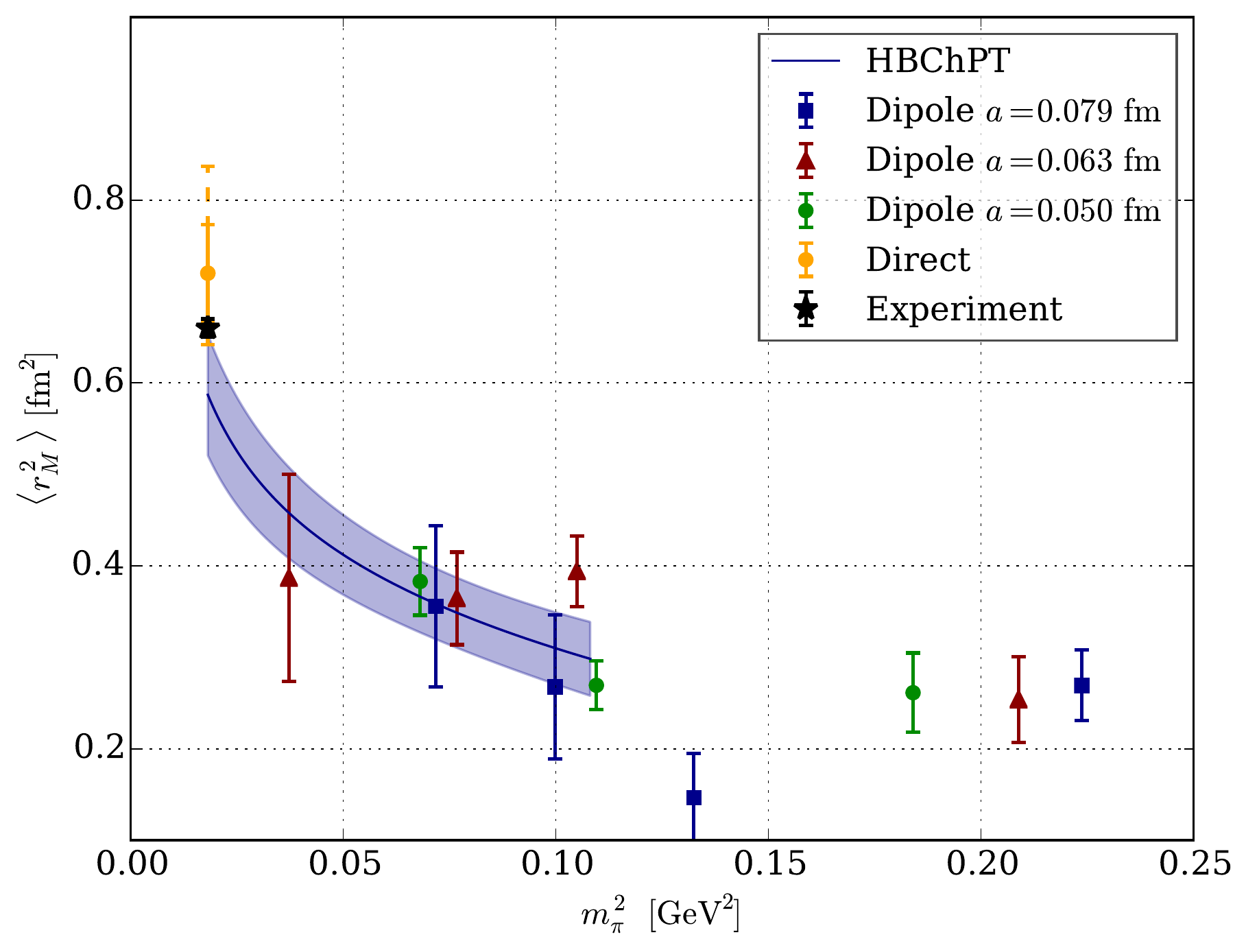}
        \caption{Pion mass dependence of electric and magnetic radii
          extracted by fitting the $Q^2$-dependence of form factors to
          a dipole form. Chiral fits to HBChPT for $m_\pi\leq330$\,MeV
          and their statistical uncertainty are represented by the
          bands. The yellow points denote the results obtained by
          directly fitting the form factors to the EFT
          expressions. The associated systematic uncertainties are
          shown by the dashed error bars. \label{fig:chiral}}
        \includegraphics[width=.46\linewidth]{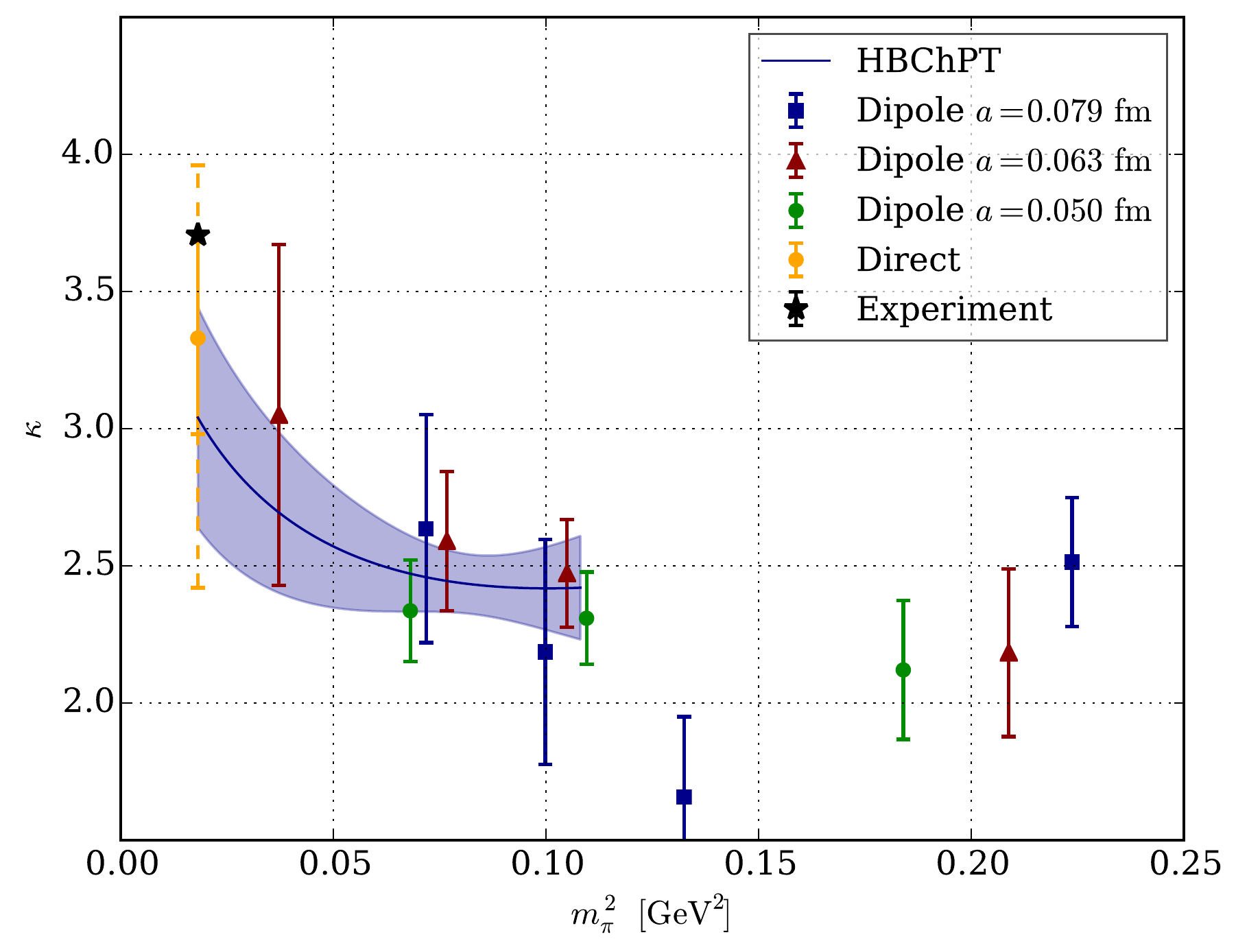}
        \caption{Pion mass dependence of the anomalous magnetic
          moment. For an explanation of symbols, see
          Fig.\,\ref{fig:chiral}. \label{fig:chiral2}}
\end{figure*}

In Figs.\,\ref{fig:chiral} and\,\ref{fig:chiral2} we compare the
estimates for $\langle r_{\rm E}^2\rangle$, $\langle r_{\rm
  M}^2\rangle$ and $\kappa$ at the physical point (shown as yellow
points) with experiment. While $\langle r_{\rm M}^2\rangle$ and
$\kappa$ agree quite well with the experimental results within
statistical errors, we find that direct fits to the form factors
underestimate the electric radius. However, given the large
systematic uncertainty, we note that our estimate for $\langle r_{\rm
  E}^2\rangle$ is not incompatible with either the CODATA
result\,\cite{Mohr:2012tt} or the value determined from muonic
hydrogen\,\cite{Pohl:2010zza,Antognini:1900ns}.

In order to enable a comparison with previous lattice determinations
\cite{Alexandrou:2006ru,nuclFF:RBC08_nf2,Yamazaki:2009zq,
  Syritsyn:2009mx,nuclFF:QCDSF_lat09,Bratt:2010jn,nuclFF:QCDSF_lat10,
  Alexandrou:2011db,Gockeler:2011ze,Collins:2011mk,Green:2013hja,
  Jager:2013kha,Green:2014xba}, we have performed chiral
extrapolations of our data to the expressions of HBChPT including the
$\Delta$-resonance at $\mathcal{O}(\epsilon^3)$ in the small scale
expansion\,\cite{Bernard:1998gv}. In particular, we fitted the pion
mass dependence of our data for the Dirac radius $\langle
r_1^2\rangle$, the anomalous magnetic moment $\kappa$ and the
combination $\kappa\langle r_2^2\rangle$ to the expressions of
refs.\,\cite{Bernard:1998gv,Hemmert:2002uh,Gockeler:2003ay}, which are
summarized, e.g. in appendix~A of\,\cite{Green:2014xba}. In these fits
the low-energy parameters $g_{\rm A}$, $F_\pi$ and $g_{\pi N\Delta}$
have been fixed to the same values as in Table\,\ref{tab:VChPT}. An
additional parameter, the photon-nucleon-$\Delta$ coupling, was fixed
to the value $-2.26$\,\cite{Hemmert:2002uh}.

The results of such a HBChPT fit, with a pion mass cut of $330$\,MeV,
without terms parameterizing cutoff effects and with the masses of the
nucleon and $\Delta$ fixed to $m_{\rm N}=938$\,MeV and
$m_\Delta=1210$\,MeV, respectively, are shown in
Fig.\,\ref{fig:chiral}. While the value for $\langle r_{\rm
  M}^2\rangle$ agrees within statistical errors with the result
determined from directly fitting the form factors, there is a
deviation by more than two standard deviations in the case of $\langle
r_{\rm E}^2\rangle$, which, however, becomes insignificant when
systematic errors are taken into account. Interestingly, the result
for the electric radius obtained from the HBChPT fit is statistically
compatible with both the CODATA estimate and the value determined from
muonic hydrogen. However, a number of comments are in order: firstly,
we note that HBChPT fits including terms parameterizing lattice
artefacts mostly fail. This may be due to the lack of a clear trend in
the data for the charge radii as the lattice spacing is
varied. Secondly, fits based on HBChPT depend much more strongly on
whether the input data originate from applying the summation method or
two-state fits.

We note in passing that we have also applied baryonic EFT based on the
Lagrangian described in Table\,\ref{tab:VChPT} to perform chiral
extrapolations of charge radii and $\kappa$ as an alternative to
HBChPT. The results at the physical point are consistent with direct
fits to the form factors within statistical errors, except for the
anomalous magnetic moment. We conclude that the chiral behaviour of
the quantities computed here is not fully understood in terms of
baryonic chiral effective theory. In order to clarify the situation it
is mandatory to obtain more statistically precise data at the physical
pion mass.

Given that the results for the charge radii and $\kappa$ exhibit a
large spread depending on the details of the chiral fitting procedure,
the agreement of the HBChPT result for $\langle r_{\rm E}^2\rangle$
with the experimental values must be considered an accident. Due to
the better overall stability we prefer to quote our final estimates
from fits applied directly to the form factors $\GX{E}$ and $\GX{M}$
(see eq.\,(\ref{eq:bestresults})).

\section{Conclusions and outlook\label{conclusions}}

We have performed a comprehensive study of the iso-vector
electromagnetic form factors of the nucleon in two-flavour QCD with a
strong emphasis on controlling the various sources of systematic
error. Our findings culminate in the estimates shown in
eq.\,(\ref{eq:bestresults}) for the charge radii and magnetic moment
including a full error budget.

While the summation method provides a handle to explore excited-state
contributions independently of the standard plateau method, the issue
of a systematic bias could not be fully resolved. Although we prefer
the summation method, since it does not rely on specific values of the
energy gaps, two-state fits produce values that appear to reproduce
the phenomenological situation better. However, for lack of data at
source-sink separations of more than $1.5$\,fm, we cannot currently
resolve the issue completely. Still, our data support the notion that
agreement with experiment can be obtained by mapping out the pion mass
dependence close to the physical point in conjunction with addressing
the issue of excited-state contaminations, which has a greater impact
as the pion mass is reduced.

For the first time, we publish a complete error budget for baryonic
charge radii and magnetic moments. Also for the first time, we have
applied the method of~%
\cite{Bauer:2012pv},
i.e. applying the full framework of covariant baryonic chiral
effective theory to the nucleon electromagnetic form factors, in the
context of lattice QCD.

While the wider picture of the convergence properties of the various
forms of baryonic chiral effective theory cannot be fully addressed
with our data, we have a clear preference for applying the fully
covariant chiral effective theory to the form factors themselves. In
particular, this avoids the systematic uncertainties inherent in the
use of the somewhat simplistic dipole description of the form factors.

In order to further resolve the issue of excited-state effects, significant
improvements in statistical precision at larger source-sink separations
will be required. One proposed way to achieve this is the use of methods
such as All-Mode Averaging (AMA)~%
\cite{Shintani:2014vja}
in order to reduce the associated computational costs. First studies
have yielded encouraging results, and we intend to pursue this avenue
further. Another direction for improvement will be to make use of the
variational method
\cite{Michael:1982gb,Luscher:1990ck,Blossier:2009kd,Bulava:2011yz}
for increased control and suppression of excited-state contributions.

Going beyond the iso-vector form factors to the proton and neutron form
factors will require the inclusion of quark-disconnected diagrams, which
makes it necessary to use all-to-all propagators. By employing stochastic
estimators along with a generalized hopping parameter expansion (gHPE),
the scalar charge radius of the pion, which receives a significant
contribution from quark-disconnected diagrams, has recently been measured
on the lattice with an overall accuracy rivalling that of phenomenological
determinations
\cite{Gulpers:2013uca,Guelpers:2014vra,Guelpers:PhD}.
Related methods have been employed by other groups to study the
nucleon structure quantities with disconnected parts
\cite{Doi:2009sq,Babich:2010at,Abdel-Rehim:2013wlz,Gupta:2015tpa},
and we intend to further address the proton radius puzzle from the lattice
side by using these and similar methods to separately study proton and
neutron form factors in the future.


\begin{acknowledgments}
The authors acknowledge the contributions of Andreas J\"uttner to the
early stages of this work. We thank Jeremy Green for useful
discussions concerning the results of
refs.\,\cite{Green:2012ud,Green:2014xba}, and Eigo Shintani for
discussions on All-Mode Averaging. We are grateful to our colleagues
within the CLS initiative for sharing ensembles. These calculations
were partly performed on the HPC Cluster ``Wilson'' at the Institute
for Nuclear Physics, University of Mainz. We thank Christian Seiwerth
for technical support. We are grateful for computer time allocated to
project HMZ21 on the BG/Q ``JUQUEEN'' computer at NIC, J\"ulich. This
work was granted access to the HPC resources of the Gauss Center for
Supercomputing at Forschungzentrum J\"ulich, Germany, made available
within the Distributed European Computing Initiative by the PRACE-2IP,
receiving funding from the European Community's Seventh Framework
Programme (FP7/2007-2013) under grant agreement RI-283493. This work
was supported by the DFG through SFB\,443 and SFB\,1044, and by the
Rhineland-Palatinate Research Initiative. MDM was partially supported
by the Danish National Research Foundation under grant number DNRF:90.
TR was supported by DFG grant HA4470/3-1.  HW is grateful to the
Yukawa Institute for Theoretical Physics, Kyoto University for
hospitality during the YITP workshop YITP-T-14-03 on ``Hadrons and
Hadron Interactions in QCD'' where part of this work was completed.
GvH thanks the Tata Institute for Fundamental Research (Mumbai, India)
for its hospitality during the workshop ``Perspectives and Challenges
in Lattice Gauge Theories'', February 2015.
\end{acknowledgments}


\appendix
\section{On the use of non-covariant interpolating operators in nucleon form factor calculations\label{app_smear}}

Typically, the interpolating nucleon operators used in lattice form-factor calculations 
are not Lorentz covariant, due to the spatial smoothing procedures applied on the elementary fields.
If we denote the overlap of the interpolating operator $\Psi_\alpha(x)$ onto the nucleon 
as follows\footnote{The states are normalized according to
$
\langle N,\vec p',s'|N,\vec p,s\rangle  = 2E_{\vec p}\,\delta_{ss'}\,(2\pi)^3 \delta^{(3)}(\vec p'-\vec p).
$
},
\begin{eqnarray}\label{eq:ovs}
\langle 0|\Psi_\alpha(0,\vec x)|N,\vec p,s\rangle  &=&   U_\alpha^s(\vec p)\,e^{i\vec p\cdot\vec x},
\end{eqnarray}
the question then poses itself, whether the fact that $U_\alpha^s(\vec p)$ does not obey the 
Dirac equation affects the calculation in any way. 

We denote by $u^s(\vec p)$ 
the usual plane-wave solutions to the Dirac equation\footnote{In this appendix we use the conventions of 
Peskin and Schroeder in Minkowski space (with Dirac matrices $\gamma^\mu_{^M}$) 
and introduce a corresponding set of Euclidean Dirac matrices,
$\gamma_0 = \gamma^0_{^M}$ and $\gamma_k = -i\gamma^k_{^M}$. The $\gamma_\mu$ are all hermitian; $\gamma_0$, $\gamma_2$
and $\gamma_5$ are symmetric, $\gamma_1$ and $\gamma_3$ are antisymmetric.}, 
norma\-lized according to 
\begin{equation}
\bar u^r(\vec p) u^s(\vec p) = 2m_N\delta^{rs}
\end{equation}
and obeying the spin sum rule
\begin{equation}\label{eq:Essr}
\sum_{s=1,2} u^s(\vec p) \bar u^{s}(\vec p) = E_{\vec p}\gamma_0 -i\vec p\cdot\vec\gamma+ m_N.
\end{equation}
Let $\psi_\alpha(x)$ be a nucleon interpolating field which does transform as a covariant Dirac spinor.
Then its overlap onto the nucleon state has the form
\begin{equation}
\langle 0|\psi_\alpha(0,\vec x)|N,\vec p,s\rangle  = Z_l \, u_\alpha^s(\vec p)\, e^{i\vec p\cdot\vec x} ,
\end{equation}
where $Z_l$ is independent of $\vec p$ and can be chosen real and positive by an appropriate choice of 
the phase of the field $\psi_\alpha(x)$.  To answer the question
formulated above, we analyze the covariance properties of correlation
functions of the operators $\Psi_\alpha(x)$ and $\psi_\alpha(x)$.
We will focus on the asymptotic behaviour of the correlation functions at large 
Euclidean time separations, where they are saturated by the nucleon ground state, 
and we indicate this by a $\star$ in the equations below.

We consider the following two-point functions and their spectral representation for $x_0>0$,
\begin{eqnarray}
\label{eq:C2ss}
C^{\alpha\beta}_{2,ss}(\vec p,x_0) &\equiv & 
\int d^3x\; e^{-i\vec p\cdot\vec x} \;\langle \Psi_\alpha(x)\bar\Psi_\beta(0)\rangle \nonumber\\
&\stackrel{\star}{=}& \frac{e^{-E_{\vec p}x_0}}{2E_{\vec p}} 
 \sum_{s}   U_\alpha^{s}(\vec p) \bar U_\beta^{s}(\vec p),
\\
C_{2,sl}^{\alpha\beta}(\vec p,x_0) &\equiv &
\int d^3x\; e^{-i\vec p\cdot\vec x} \;\langle \Psi_\alpha(x)\bar\psi_\beta(0)\rangle \nonumber\\
&\stackrel{\star}{=}&
  Z_l^*\;\frac{e^{-E_{\vec p}x_0}}{2E_{\vec p}}\; \sum_{s}  U_\alpha^{s}(\vec p) \bar u_\beta^{s}(\vec p), 
\label{eq:C2sl}
\\
C_{2,ls}^{\alpha\beta}(\vec p,x_0) &\equiv &
\int d^3x\; e^{-i\vec p\cdot\vec x} \;\langle \psi_\alpha(x)\bar\Psi_\beta(0)\rangle \nonumber\\
&\stackrel{\star}{=}& 
  Z_l\;\frac{e^{-E_{\vec p}x_0}}{2E_{\vec p}}\; \sum_{s} u_\alpha^s(\vec p)  \bar U_\beta^s(\vec p).
\label{eq:C2ls}
\end{eqnarray}
We now define
\begin{equation}\label{eq:calMdef}
{\cal M}(\vec p) \equiv  \sum_{s}  U^{s}(\vec p) \bar u^{s}(\vec p).
\end{equation}
The other spin sums can also be expressed through ${\cal M}$. First, we have 
\begin{equation}
\sum_{s} u^s(\vec p)  \bar U^s(\vec p) = \gamma_0 {\cal M}(\vec p)^\dagger \gamma_0.
\end{equation}
Since the overlap of the local nucleon operator is a genuine Dirac spinor, the matrix ${\cal M}$ satisfies
\begin{equation}\label{eq:MpD}
{\cal M}(\vec p) (\gamma_0 E_{\vec p} -i\vec p\cdot\vec \gamma - m_N) =0.
\end{equation}
We observe from the definition of ${\cal M}$  that 
\begin{equation}
{\cal M}(\vec p) u^s(\vec p) = 2m_N U^s(\vec p),
\end{equation}
and hence also
\begin{equation}
\bar U^s(\vec p) = \frac{1}{2m_N} \bar u^s(\vec p) \gamma_0 {\cal M}(\vec p)^\dagger \gamma_0.
\end{equation}
We can thus write the spin sum appearing in the two-point function $C_{2,ss}$ as 
\begin{eqnarray}\label{eq:UUbar}
 \sum_{s}  U^{s}(\vec p) \bar U^{s}(\vec p)
&=& \frac{1}{4m_N^2} {\cal M}(\vec p) (E_{\vec p}\gamma_0 -i\vec p\cdot\vec\gamma + m_N) \nonumber\\
&& \times~\gamma_0 {\cal M}(\vec p)^\dagger \gamma_0
\\ &=& 
\frac{1}{2m_N} {\cal M}(\vec p) \gamma_0 {\cal M}(\vec p)^\dagger \gamma_0.
\end{eqnarray}
The second equality uses Eq.~(\ref{eq:MpD}).
Thus $\gamma_0$ times the spin sum of the $U$'s is a hermitian matrix.

Let $J(x)$ be a local operator with the following matrix elements between one-nucleon 
states\footnote{For instance,
${\cal J}(\vec q) = \left(\gamma_{^M}^\mu F_1(q^2) + i\sigma_{^M}^{\mu\nu}\frac{q_\nu}{2m_N} F_2(q^2)\right)$ 
($\sigma^{\mu\nu}_{^M}\equiv \frac{i}{2}[\gamma^\mu_{^M},\gamma^\nu_{^M}]$)
for the vector current $\bar\psi(x) \gamma_{^M}^\mu \psi(x)$.},
\begin{equation}\label{eq:defcalJ}
\langle N,\vec p',s'| J(0)|N,\vec p,s\rangle  = \bar u^{s'}(\vec p') {\cal J}(\vec q) u^s(\vec p),
\end{equation}
where $\vec q\equiv \vec p'-\vec p$.
The three-point function relevant to form factor calculations with vanishing momentum at the sink
and its  spectral representation read (for $x_0>y_0>0$)
\begin{eqnarray}
C^{\alpha\beta}_{3, J}(\vec q,y_0,x_0)&\equiv& \int d^3y\int d^3x \; e^{+i\vec q\cdot\vec y}
\langle \Psi_\alpha(x)\; J(y)\;\overline{\Psi}_\beta(0)\rangle \nonumber\\
&\stackrel{\star}{=} &   \frac{e^{-E_{\vec q}y_0}}{2E_{\vec q}}\;\frac{e^{-m(x_0-y_0)}}{2m_N}
\Big( \sum_{s'}  U_\alpha^{s'}(\vec 0) \bar u_\gamma^{s'}(\vec 0)  \Big)\nonumber\\
&\times& \big({\cal J}(\vec q)\big)_{\gamma\delta} 
 \Big(\sum_{s} u_{\delta}^s(-\vec q)  \bar U_\beta^s(-\vec q)\Big).
\end{eqnarray}
The three-point function, projected with a generic projector $\Gamma$, 
\begin{eqnarray}\label{eq:C3ptgen}
{\rm Tr\,}\big\{\Gamma C_{3}(\vec q,y_0,x_0)\big\}  &\stackrel{\star}{=}&
\frac{e^{-E_{\vec q}y_0}}{2E_{\vec q}}\;\frac{e^{-m(x_0-y_0)}}{2m_N}\\
&\times&{\rm Tr}\Big\{\Gamma {\cal M}(\vec 0)  {\cal J}(\vec q) \gamma_0 {\cal M}(-\vec q)^\dagger \gamma_0\Big\}\nonumber
\end{eqnarray}
can thus also be expressed in terms of  the matrix ${\cal M}(\vec p)$. 
The latter can be expanded in the 16 linearly independent spinor-space matrices, 
and symmetries can be used to restrict the terms that contribute.

\subsection{Symmetry constraints on nucleon two-point functions}

Let
\begin{equation}
C_{2,\phi\chi}(\vec p,x_0) \equiv  \int d^3x \;e^{-i\vec p\cdot\vec x}\; \big\langle \phi(x)\, \bar\chi(0)\big\rangle 
\end{equation}
be a generic nucleon two-point function with interpolating operators $\phi$ and $\chi$.
We assume that the latter are good spinors with respect to spatial rotations
and with respect to all discrete symmetries, but no assumption is made on their behaviour 
under boosts. The implications of the various symmetries are:
\begin{itemize}
\item rotation invariance:
\begin{equation}
C_{2,\phi\chi}(R\vec p,x_0) = \Lambda_{\frac{1}{2}}(R) C_{2,\phi\chi}(\vec p,x_0) \Lambda_{\frac{1}{2}}(R)^{-1},
\end{equation}
where $\phi (x) \to \Lambda_{\frac{1}{2}}(R) \phi(R^{-1} x)$ is the transformation law of the quark fields;
\item parity:
\begin{equation}
C_{2,\phi\chi}(-\vec p,x_0) = \gamma_0 C_{2,\phi\chi}(\vec p,x_0)\gamma_0\;;
\end{equation}
\item Euclidean time reversal:
\begin{equation}
C_{2,\phi\chi}(\vec p,-x_0) = \gamma_0\gamma_5\; C_{2,\phi\chi}(\vec p,x_0) \;\gamma_5\gamma_0\;;
\end{equation}
\item charge conjugation: assuming that the nucleon interpolating fields transform like the quark fields
($q_\alpha(x) \to (\bar q(x) \gamma_0\gamma_2)_\alpha $,
$\bar q_\alpha(x)\to (\gamma_0\gamma_2 q(x))_\alpha$)\footnote{This condition fixes the phase of the interpolating fields.
An interpolating field that satisfies this condition is $\chi_\alpha(x)=
\epsilon^{abc} (u^a_\beta (\gamma_0\gamma_2\gamma_5)_{\beta\gamma} d^b_\gamma) u^c_\alpha $,
but $e^{i\varphi}\chi_\alpha(x)$ does not for $\varphi\neq 0,\pi$.}
\begin{equation}
C_{2,\phi\chi}(-\vec p,-x_0)^\top = \gamma_2\gamma_0 \,C_{2,\chi\phi}(\vec p,x_0) \gamma_0\gamma_2.
\end{equation}
\end{itemize}
Combining the three discrete symmetries, we obtain, for later use,
\begin{equation}\label{eq:CPT}
\gamma_5 C_{2,\phi\chi}(\vec p,x_0)^\top \gamma_5 = 
\gamma_2 \gamma_0 C_{2,\chi\phi}(\vec p,x_0) \gamma_0\gamma_2.
\end{equation}

The most general form of the nucleon two-point functions
allowed by rotation symmetry and parity is
\begin{eqnarray}\label{eq:C2psichi}
C_{2,\phi\chi}(\vec p,x_0) &=&{\cal F}_s(\vec p^2,x_0) 
+  {\cal F}_0(\vec p^2,x_0) \gamma_0 \\
&&-i {\cal F}_V(\vec p^2,x_0) \vec p\cdot\vec\gamma 
-i {\cal F}_T(\vec p^2,x_0) \gamma_0\;\vec p\cdot\vec\gamma.\nonumber
\end{eqnarray}
Time-reversal invariance implies that ${\cal F}_0$ and ${\cal F}_T$ are odd functions of $x_0$,
while ${\cal F}_s$ and ${\cal F}_V$ are even functions of $x_0$.
Charge conjugation relates the functions ${\cal F}$ parametrizing the correlator $C_{2,\phi\chi}$ to those 
parametrizing the correlator $C_{2,\chi\phi}$.

\subsection{General parametrization of  ${\cal M}(\vec p)$}

In view of (\ref{eq:C2psichi}), we parametrize ${\cal M}(\vec p)$ as 
\begin{eqnarray}
{\cal M}(\vec p) &=& 
m_N f_s(\vec p^2) + f_0(\vec p^2) E_{\vec p}\gamma_0  -i f_V(\vec p^2) \vec p\cdot\vec\gamma\nonumber\\
&&~-i f_T(\vec p^2) \vec p \cdot (\gamma_0 \vec\gamma).
\end{eqnarray}
The condition (\ref{eq:MpD}) from the spectral representation implies the constraints
\begin{eqnarray}\label{eq:c1}
f_V(\vec p^2) &=& \frac{1}{\vec p^2}\Big( E_{\vec p}^2 \, f_0(\vec p^2) - m_N^2 f_s(\vec p^2)\Big),
\\
f_T(\vec p^2) &=& \frac{m_N \, E_{\vec p}}{\vec p^2} \Big(f_s(\vec p^2) - f_0(\vec p^2)\Big).
\label{eq:c2}
\end{eqnarray}
In particular, $ f_0(\vec p^2) -f_s(\vec p^2) =  {\rm O}(\vec p^2)$;
the case of a Lorentz-covariant spinor source corresponds to 
$f_0=f_s=f_V$ independent of $\vec p$ and $f_T=0$ identically.
Now, combining the discrete-symmetry property (\ref{eq:CPT}) and the
spectral representations (\ref{eq:C2sl}, \ref{eq:C2ls}), one derives
the property that $f_0$, $f_s$, $f_V$ and $f_T$ must all be real-valued functions.

It is convenient to decompose ${\cal M}$ as follows,
\begin{eqnarray}
{\cal M}(\vec p) &=& {\cal M}_+(\vec p) + {\cal M}_-(\vec p),  \\
{\cal M}_\pm(\vec p) &=& {\textstyle\frac{1}{2}} (1\pm\gamma_0) {\cal M}(\vec p).
\end{eqnarray}
Using the constraints, one finds the following general form of ${\cal M}_\pm(\vec p)$,
\begin{eqnarray}\label{eq:Mproj}
{\cal M}_+(\vec p) &=& {\textstyle\frac{1}{2}}(1+\gamma_0) \; Z_+(\vec p) (E_{\vec p} + m_N -i\vec p\cdot\vec \gamma),\quad
\\
  {\cal M}_-(\vec p) &=& {\textstyle\frac{1}{2}}(1-\gamma_0) \; Z_-(\vec p) ( E_{\vec p} - m_N +i\vec p\cdot\vec \gamma)\quad
\end{eqnarray}
with the relations
\begin{eqnarray}
Z_+(\vec p) =  \frac{m_Nf_s(\vec p^2) + E_{\vec p}f_0(\vec p^2)}{E_{\vec p} + m_N},\\
Z_-(\vec p) =  \frac{m_Nf_s(\vec p^2) - E_{\vec p}f_0(\vec p^2)}{E_{\vec p} - m_N}.
\end{eqnarray}
The bottom line is that the matrix ${\cal M}(\vec p)$ is parametrized
by two (spatially scalar) functions $Z_\pm(\vec p)$, which are linear functions of the nucleon interpolating 
operator $\Psi_\alpha(x)$. Its phase can be chosen such that $Z_\pm(\vec p)$ 
are real for all $\vec p$. Furthermore it can be chosen such that $Z_+(\vec p=0)$ is positive.
The generic case that we will consider is then that it remains positive for all momenta\footnote{A priori, it could 
happen that it becomes negative for some momenta, but then by continuity 
it would have to vanish somewhere, and there is no symmetry reason for this to happen.}.
As a side remark, we note that in the rest frame only parity-odd states (not considered here)
contribute to ${\rm Tr\,}\{(1-\gamma_0)C_{2,sl}(\vec p=0,x_0)\}$.

\subsection{Final form of the non-covariant two-point and three-point functions}

We now specialize to the projector
\begin{equation}\label{eq:Gp}
\Gamma = {\textstyle\frac{1}{2}}(1+\gamma_0) (1+i\gamma_5\gamma_3).
\end{equation}
Starting from Eq.~(\ref{eq:C2ss}) and (\ref{eq:UUbar}), the projected two-point function can 
be written\footnote{In the following equations
we allow for the case that the $Z_+(\vec p)$ have a common
phase for all $\vec p$.}, 
\begin{eqnarray}
{\rm Tr\,} \{\Gamma C_{2,ss}(\vec p,x_0)\}
&\stackrel{\star}{=}& \frac{e^{-E_{\vec p}x_0} }{2E_{\vec p}}\cdot\frac{1}{2m_N}
{\rm Tr\,}\{ (1+i\gamma_5\gamma_3) \nonumber\\
&&~\times {\cal M}_+(\vec p) \gamma_0 {\cal M}_+(\vec p)^\dagger \gamma_0 \}
\\ &=&  |Z_+(\vec p)|^2  (1+m_N/E_{\vec p})\;e^{-E_{\vec p}x_0}.\nonumber
\end{eqnarray}
The second equality uses the parametrization Eq.~(\ref{eq:Mproj}).
The term $i\gamma_5\gamma_3$ in $\Gamma_p$ does not contribute to this expression.
The trace appearing in Eq.~(\ref{eq:C3ptgen}) takes the form
\begin{eqnarray}
& {\rm Tr}&\Big\{\Gamma_p {\cal M}(\vec 0)  {\cal J}(\vec q) \gamma_0 {\cal M}(-\vec q)^\dagger \gamma_0\Big\} 
=\\
&&~{\rm Tr\,}\Big\{  (1+i\gamma_5\gamma_3) {\cal M}_+(\vec 0)  {\cal J}(\vec q) \gamma_0 
{\cal M}_+(-\vec q)^\dagger  \Big\}. \nonumber
\end{eqnarray}

Thus the expression for the three-point function becomes
\begin{eqnarray}\label{eq:3ptfinal}
&C&_{3}(\vec q,y_0,x_0) \stackrel{\star}{=} 
 \frac{e^{-E_{\vec q}y_0}}{2E_{\vec q}}\;\frac{e^{-m_N(x_0-y_0)}}{2m_N}
 Z_+(0)  Z_+(-\vec q)^*
\nonumber\\ &&  m_N\, 
{\rm Tr\,}\Big\{ (1+\gamma_0) (1+i\gamma_5\gamma_3) {\cal J}(\vec q) (m_N+E_{\vec q} + i\vec q\cdot\vec \gamma) \Big\}.
\nonumber\\
\end{eqnarray}

As we have seen, the phase of the nucleon interpolating operator $\Psi(x)$ can be chosen
such that $Z_+(\vec p)>0$ for all $\vec p$. If the phase had not been chosen in this way,
the common phase would nonetheless cancel in the product $ Z_+(0)  Z_+(-\vec q)^*$.

For a covariant source, the result would be identical to (\ref{eq:3ptfinal}), except that
$Z_+(\vec p)$ would be independent of $\vec p$.  In the standard
expression of the ratio (\ref{eq:Ratio}),
the three-point function is however divided by the
appropriate combination of two-point functions to cancel the overlap 
factor $Z_+(\vec p)$ for each value of $\vec p $. The correctness of our 
calculation is thus not affected by the use of non-covariant interpolating 
operators.

\section{$Q^2$ Tables\label{wii}}

In Tables~%
\ref{tab:Q2_A3}--\ref{tab:Q2_O7}
we give all of our results for the iso-vector vector form factors $\GX{E}$
and $\GX{M}$ of the nucleon at all values of $Q^2$ measured on each ensemble.
Listed in each case are the values obtained using the plateau method with
a source-sink separation of $t_s=1.1$\,fm, the summation method, and an
explicit two-state fit (cf. the main text for details).
The statistical errors on each data point are quoted in parentheses following
the central value.


\renewcommand\arraystretch{1.2}

\begin{table*}[h]
	\begin{center}
		\begin{tabular}{|c|c|c|c|c|c|c|}
    			\hline  
        				A3	& \multicolumn{3}{c|}{$G_E$} & \multicolumn{3}{c|}{$G_M$} \\
        			\hline
     			$Q^2$ $[\mathrm{GeV}^2]$   	& Plat (1.1 fm) & Summation & Two-state & Plat (1.1 fm) & Summation & Two-state\\
   			 \hline        
0.230 & 0.752 (0.011) & 0.725 (0.018) & 0.734 (0.012) & 3.040 (0.100) & 3.320 (0.175) & 3.067 (0.128) \\ 
0.443 & 0.601 (0.013) & 0.583 (0.025) & 0.592 (0.017) & 2.359 (0.078) & 2.503 (0.147) & 2.272 (0.111) \\ 
0.639 & 0.495 (0.018) & 0.480 (0.037) & 0.484 (0.022) & 1.921 (0.084) & 1.869 (0.176) & 1.783 (0.125) \\ 
0.823 & 0.373 (0.023) & 0.322 (0.056) & 0.369 (0.036) & 1.579 (0.098) & 1.716 (0.216) & 1.536 (0.158) \\ 
0.995 & 0.331 (0.023) & 0.299 (0.055) & 0.320 (0.030) & 1.349 (0.091) & 1.416 (0.205) & 1.253 (0.143) \\ 
1.156 & 0.283 (0.031) & 0.228 (0.073) & 0.213 (0.037) & 1.194 (0.121) & 0.881 (0.295) & 0.910 (0.162) \\   
    			\hline  
		\end{tabular}
	\end{center}
\caption{A3 ensemble ($a=0.079$\,fm, $m_\pi=473$\,MeV): Vector form factors at all $Q^2$ values for all extraction methods. \label{tab:Q2_A3}}
\end{table*}


\begin{table*}[h]
	\begin{center}
		\begin{tabular}{|c|c|c|c|c|c|c|}
    			\hline  
        				A4	& \multicolumn{3}{c|}{$G_E$} & \multicolumn{3}{c|}{$G_M$} \\
        			\hline
     			$Q^2$ $[\mathrm{GeV}^2]$   	& Plat (1.1 fm) & Summation & Two-state & Plat (1.1 fm) & Summation & Two-state\\
   			 \hline        
0.229 & 0.720 (0.024) & 0.685 (0.037) & 0.668 (0.021) & 2.806 (0.147) & 2.603 (0.232) & 2.726 (0.239) \\ 
0.437 & 0.547 (0.022) & 0.466 (0.043) & 0.489 (0.031) & 2.121 (0.109) & 2.360 (0.210) & 2.518 (0.210) \\ 
0.628 & 0.459 (0.024) & 0.380 (0.048) & 0.385 (0.045) & 1.936 (0.130) & 2.166 (0.199) & 2.146 (0.246) \\ 
0.805 & 0.401 (0.039) & 0.397 (0.087) & 0.339 (0.076) & 1.591 (0.239) & 1.499 (0.339) & 1.408 (0.375) \\ 
0.970 & 0.338 (0.025) & 0.338 (0.080) & 0.291 (0.052) & 1.317 (0.128) & 1.449 (0.281) & 1.594 (0.262) \\ 
1.123 & 0.299 (0.035) & 0.242 (0.099) & 0.205 (0.077) & 1.046 (0.156) & 1.192 (0.352) & 1.510 (0.409) \\    
    			\hline  
		\end{tabular}
	\end{center}
\caption{A4 ensemble ($a=0.079$\,fm, $m_\pi=364$\,MeV): Vector form factors at all $Q^2$ values for all extraction methods. \label{tab:Q2_A4}}
\end{table*}


\begin{table*}[h]
	\begin{center}                                                           
		\begin{tabular}{|c|c|c|c|c|c|c|}
    			\hline  
        				A5	& \multicolumn{3}{c|}{$G_E$} & \multicolumn{3}{c|}{$G_M$} \\
        			\hline
     			$Q^2$ $[\mathrm{GeV}^2]$   	& Plat (1.1 fm) & Summation & Two-state & Plat (1.1 fm) & Summation & Two-state\\
   			 \hline        
0.228 & 0.692 (0.018) & 0.651 (0.036) & 0.629 (0.039) & 2.572 (0.159) & 2.619 (0.270) & 2.672 (0.349) \\ 
0.434 & 0.521 (0.019) & 0.461 (0.038) & 0.433 (0.040) & 1.967 (0.121) & 2.082 (0.203) & 2.021 (0.272) \\ 
0.623 & 0.424 (0.029) & 0.305 (0.058) & 0.328 (0.054) & 1.720 (0.134) & 2.115 (0.277) & 1.408 (0.328) \\ 
0.797 & 0.343 (0.043) & 0.269 (0.096) & 0.171 (0.087) & 1.247 (0.158) & 0.673 (0.426) & 1.430 (0.442) \\ 
0.959 & 0.255 (0.033) & 0.176 (0.084) & 0.002 (0.075) & 1.023 (0.135) & 0.989 (0.350) & 0.867 (0.381) \\ 
1.110 & 0.065 (0.093) & 0.189 (0.114) & 0.067 (0.128) & 0.760 (0.340) & 0.750 (0.523) & -0.338 (0.522) \\       
    			\hline  
		\end{tabular}
	\end{center}
\caption{A5 ensemble ($a=0.079$\,fm, $m_\pi=316$\,MeV): Vector form factors at all $Q^2$ values for all extraction methods. \label{tab:Q2_A5}}
\end{table*}


\begin{table*}[h]
	\begin{center}
		\begin{tabular}{|c|c|c|c|c|c|c|}
    			\hline  
        				B6	& \multicolumn{3}{c|}{$G_E$} & \multicolumn{3}{c|}{$G_M$} \\
        			\hline
     			$Q^2$ $[\mathrm{GeV}^2]$   	& Plat (1.1 fm) & Summation & Two-state & Plat (1.1 fm) & Summation & Two-state\\
   			 \hline        
0.104 & 0.826 (0.008) & 0.785 (0.023) & 0.760 (0.018) & 3.264 (0.121) & 3.692 (0.370) & 3.794 (0.284) \\ 
0.203 & 0.709 (0.012) & 0.664 (0.036) & 0.669 (0.025) & 2.819 (0.121) & 3.172 (0.258) & 2.571 (0.260) \\ 
0.297 & 0.622 (0.017) & 0.572 (0.048) & 0.541 (0.038) & 2.312 (0.121) & 2.417 (0.245) & 2.283 (0.299) \\ 
0.387 & 0.565 (0.018) & 0.569 (0.069) & 0.474 (0.042) & 2.154 (0.132) & 2.435 (0.241) & 2.115 (0.300) \\ 
0.474 & 0.507 (0.018) & 0.460 (0.053) & 0.428 (0.044) & 2.034 (0.105) & 2.149 (0.216) & 1.970 (0.247) \\ 
0.557 & 0.445 (0.020) & 0.421 (0.059) & 0.374 (0.056) & 1.656 (0.069) & 1.680 (0.208) & 1.399 (0.233) \\      
    			\hline  
		\end{tabular}
	\end{center}
\caption{B6 ensemble ($a=0.079$\,fm, $m_\pi=268$\,MeV): Vector form factors at all $Q^2$ values for all extraction methods. \label{tab:Q2_B6}}
\end{table*}


\begin{table*}[h]
	\begin{center}
		\begin{tabular}{|c|c|c|c|c|c|c|}
    			\hline  
        				E5	& \multicolumn{3}{c|}{$G_E$} & \multicolumn{3}{c|}{$G_M$} \\
        			\hline
     			$Q^2$ $[\mathrm{GeV}^2]$   	& Plat (1.1 fm) & Summation & Two-state & Plat (1.1 fm) & Summation & Two-state\\
   			 \hline        
0.356 & 0.663 (0.013) & 0.635 (0.018) & 0.601 (0.014) & 2.611 (0.105) & 2.544 (0.150) & 2.540 (0.139) \\ 
0.675 & 0.477 (0.016) & 0.443 (0.025) & 0.411 (0.018) & 1.909 (0.090) & 1.773 (0.130) & 1.782 (0.111) \\ 
0.966 & 0.379 (0.026) & 0.357 (0.039) & 0.339 (0.029) & 1.429 (0.104) & 1.346 (0.161) & 1.350 (0.151) \\ 
1.233 & 0.209 (0.077) & 0.197 (0.065) & 0.204 (0.049) & 1.342 (0.358) & 1.015 (0.242) & 1.016 (0.210) \\ 
1.480 & 0.220 (0.046) & 0.191 (0.060) & 0.190 (0.038) & 0.996 (0.178) & 0.887 (0.242) & 0.929 (0.172) \\ 
1.709 & 0.204 (0.086) & 0.164 (0.082) & 0.185 (0.054) & 0.454 (0.301) & 0.644 (0.301) & 0.876 (0.241) \\  
    			\hline  
		\end{tabular}
	\end{center}
\caption{E5 ensemble ($a=0.063$\,fm, $m_\pi=457$\,MeV): Vector form factors at all $Q^2$ values for all extraction methods. \label{tab:Q2_E5}}
\end{table*}


\begin{table*}[h]
	\begin{center}
		\begin{tabular}{|c|c|c|c|c|c|c|}
    			\hline  
        				F6	& \multicolumn{3}{c|}{$G_E$} & \multicolumn{3}{c|}{$G_M$} \\
        			\hline
     			$Q^2$ $[\mathrm{GeV}^2]$   	& Plat (1.1 fm) & Summation & Two-state & Plat (1.1 fm) & Summation & Two-state\\
   			 \hline        
0.162 & 0.766 (0.009) & 0.746 (0.011) & 0.725 (0.009) & 2.834 (0.125) & 2.963 (0.143) & 2.586 (0.216) \\ 
0.314 & 0.627 (0.012) & 0.604 (0.015) & 0.555 (0.014) & 2.410 (0.102) & 2.350 (0.116) & 2.128 (0.166) \\ 
0.457 & 0.541 (0.016) & 0.516 (0.019) & 0.455 (0.020) & 2.028 (0.103) & 1.809 (0.128) & 1.962 (0.173) \\ 
0.593 & 0.423 (0.017) & 0.394 (0.023) & 0.332 (0.027) & 1.586 (0.084) & 1.505 (0.116) & 1.474 (0.199) \\ 
0.723 & 0.383 (0.017) & 0.343 (0.024) & 0.307 (0.025) & 1.435 (0.072) & 1.245 (0.111) & 1.306 (0.165) \\ 
0.846 & 0.342 (0.021) & 0.291 (0.028) & 0.253 (0.030) & 1.278 (0.076) & 1.007 (0.133) & 1.205 (0.171) \\   
    			\hline  
		\end{tabular}
	\end{center}
\caption{F6 ensemble ($a=0.063$\,fm, $m_\pi=324$\,MeV): Vector form factors at all $Q^2$ values for all extraction methods. \label{tab:Q2_F6}}
\end{table*}


\begin{table*}[h]
	\begin{center}
		\begin{tabular}{|c|c|c|c|c|c|c|}
    			\hline  
        				F7	& \multicolumn{3}{c|}{$G_E$} & \multicolumn{3}{c|}{$G_M$} \\
        			\hline
     			$Q^2$ $[\mathrm{GeV}^2]$   	& Plat (1.1 fm) & Summation & Two-state & Plat (1.1 fm) & Summation & Two-state\\
   			 \hline        
0.162 & 0.759 (0.013) & 0.751 (0.016) & 0.723 (0.017) & 2.894 (0.167) & 3.106 (0.192) & 2.637 (0.329) \\ 
0.313 & 0.623 (0.020) & 0.612 (0.020) & 0.566 (0.025) & 2.383 (0.136) & 2.462 (0.158) & 1.893 (0.258) \\ 
0.455 & 0.512 (0.026) & 0.507 (0.027) & 0.421 (0.033) & 1.944 (0.111) & 2.065 (0.158) & 1.613 (0.294) \\ 
0.589 & 0.442 (0.026) & 0.410 (0.033) & 0.270 (0.048) & 1.706 (0.130) & 1.649 (0.175) & 1.284 (0.307) \\ 
0.717 & 0.417 (0.037) & 0.372 (0.031) & 0.289 (0.048) & 1.461 (0.107) & 1.465 (0.141) & 1.336 (0.302) \\ 
0.838 & 0.336 (0.036) & 0.308 (0.038) & 0.200 (0.051) & 1.295 (0.126) & 1.329 (0.165) & 0.911 (0.342) \\
    			\hline  
		\end{tabular}
	\end{center}
\caption{F7 ensemble ($a=0.063$\,fm, $m_\pi=277$\,MeV): Vector form factors at all $Q^2$ values for all extraction methods. \label{tab:Q2_F7}}
\end{table*}


\begin{table*}[h]
	\begin{center}
		\begin{tabular}{|c|c|c|c|c|c|c|}
    			\hline  
        				G8	& \multicolumn{3}{c|}{$G_E$} & \multicolumn{3}{c|}{$G_M$} \\
        			\hline
     			$Q^2$ $[\mathrm{GeV}^2]$   	& Plat (1.1 fm) & Summation & Two-state & Plat (1.1 fm) & Summation & Two-state\\
   			 \hline        
0.092 & 0.870 (0.013) & 0.928 (0.041) & 0.771 (0.030) & 3.367 (0.243) & 4.272 (0.618) & 4.555 (0.815) \\ 
0.181 & 0.731 (0.014) & 0.737 (0.046) & 0.563 (0.040) & 2.822 (0.160) & 3.293 (0.398) & 3.081 (0.546) \\ 
0.266 & 0.629 (0.016) & 0.591 (0.055) & 0.396 (0.060) & 2.464 (0.155) & 2.972 (0.383) & 2.879 (0.600) \\ 
0.347 & 0.558 (0.022) & 0.510 (0.071) & 0.292 (0.074) & 2.173 (0.131) & 3.088 (0.410) & 2.168 (0.568) \\ 
0.426 & 0.499 (0.018) & 0.473 (0.058) & 0.149 (0.065) & 1.902 (0.119) & 2.311 (0.315) & 1.862 (0.483) \\ 
0.503 & 0.458 (0.020) & 0.419 (0.068) & 0.048 (0.082) & 1.688 (0.130) & 1.899 (0.370) & 1.767 (0.492) \\   
    			\hline  
		\end{tabular}
	\end{center}
\caption{G8 ensemble ($a=0.063$\,fm, $m_\pi=193$\,MeV): Vector form factors at all $Q^2$ values for all extraction methods. \label{tab:Q2_G8}}
\end{table*}


\begin{table*}[h]
	\begin{center}
		\begin{tabular}{|c|c|c|c|c|c|c|}
    			\hline  
        				N5	& \multicolumn{3}{c|}{$G_E$} & \multicolumn{3}{c|}{$G_M$} \\
        			\hline
     			$Q^2$ $[\mathrm{GeV}^2]$   	& Plat (1.1 fm) & Summation & Two-state & Plat (1.1 fm) & Summation & Two-state\\
   			 \hline        
0.257 & 0.708 (0.008) & 0.678 (0.018) & 0.687 (0.008) & 2.858 (0.086) & 2.703 (0.164) & 2.768 (0.101) \\ 
0.494 & 0.551 (0.010) & 0.509 (0.023) & 0.504 (0.011) & 2.171 (0.063) & 2.082 (0.129) & 2.105 (0.083) \\ 
0.715 & 0.445 (0.014) & 0.414 (0.032) & 0.386 (0.016) & 1.763 (0.067) & 1.697 (0.158) & 1.639 (0.094) \\ 
0.922 & 0.397 (0.021) & 0.331 (0.049) & 0.299 (0.021) & 1.627 (0.090) & 1.129 (0.196) & 1.214 (0.111) \\ 
1.118 & 0.332 (0.019) & 0.275 (0.044) & 0.252 (0.018) & 1.310 (0.077) & 1.204 (0.176) & 1.028 (0.085) \\ 
1.303 & 0.278 (0.025) & 0.215 (0.061) & 0.190 (0.023) & 1.117 (0.097) & 0.939 (0.242) & 0.789 (0.103) \\   
    			\hline  
		\end{tabular}
	\end{center}
\caption{N5 ensemble ($a=0.050$\,fm, $m_\pi=429$\,MeV): Vector form factors at all $Q^2$ values for all extraction methods. \label{tab:Q2_N5}}
\end{table*}


\begin{table*}[h]
	\begin{center}
		\begin{tabular}{|c|c|c|c|c|c|c|}
    			\hline  
        				N6	& \multicolumn{3}{c|}{$G_E$} & \multicolumn{3}{c|}{$G_M$} \\
        			\hline
     			$Q^2$ $[\mathrm{GeV}^2]$   	& Plat (1.1 fm) & Summation & Two-state & Plat (1.1 fm) & Summation & Two-state\\
   			 \hline        
0.255 & 0.681 (0.011) & 0.644 (0.012) & 0.588 (0.012) & 2.618 (0.096) & 2.797 (0.107) & 2.452 (0.149) \\ 
0.487 & 0.512 (0.013) & 0.473 (0.014) & 0.373 (0.015) & 2.004 (0.071) & 2.083 (0.083) & 1.717 (0.124) \\ 
0.701 & 0.386 (0.017) & 0.344 (0.020) & 0.235 (0.025) & 1.433 (0.070) & 1.688 (0.091) & 1.236 (0.144) \\ 
0.900 & 0.363 (0.026) & 0.331 (0.031) & 0.173 (0.036) & 1.341 (0.102) & 1.754 (0.127) & 1.177 (0.202) \\ 
1.087 & 0.282 (0.018) & 0.271 (0.026) & 0.152 (0.030) & 1.041 (0.068) & 1.090 (0.101) & 0.724 (0.158) \\ 
1.264 & 0.222 (0.021) & 0.204 (0.033) & 0.120 (0.039) & 0.763 (0.071) & 0.876 (0.128) & 0.390 (0.181) \\        
    			\hline  
		\end{tabular}
	\end{center}
\caption{N6 ensemble ($a=0.050$\,fm, $m_\pi=331$\,MeV): Vector form factors at all $Q^2$ values for all extraction methods. \label{tab:Q2_N6}}
\end{table*}


\begin{table*}[h]
	\begin{center}
		\begin{tabular}{|c|c|c|c|c|c|c|}
    			\hline  
        				O7	& \multicolumn{3}{c|}{$G_E$} & \multicolumn{3}{c|}{$G_M$} \\
        			\hline
     			$Q^2$ $[\mathrm{GeV}^2]$   	& Plat (1.1 fm) & Summation & Two-state & Plat (1.1 fm) & Summation & Two-state\\
   			 \hline        
0.146 & 0.786 (0.010) & 0.773 (0.012) & 0.700 (0.017) & 2.800 (0.126) & 2.825 (0.139) & 2.886 (0.259) \\ 
0.282 & 0.648 (0.013) & 0.618 (0.015) & 0.485 (0.021) & 2.230 (0.096) & 2.299 (0.118) & 1.976 (0.205) \\ 
0.411 & 0.554 (0.015) & 0.521 (0.019) & 0.357 (0.026) & 1.929 (0.086) & 1.964 (0.111) & 1.449 (0.211) \\ 
0.533 & 0.474 (0.018) & 0.401 (0.022) & 0.231 (0.035) & 1.743 (0.100) & 1.589 (0.111) & 1.059 (0.239) \\ 
0.650 & 0.428 (0.017) & 0.351 (0.023) & 0.172 (0.033) & 1.445 (0.075) & 1.387 (0.096) & 0.755 (0.194) \\ 
0.761 & 0.388 (0.019) & 0.327 (0.029) & 0.176 (0.033) & 1.249 (0.070) & 1.266 (0.096) & 0.662 (0.210) \\       
    			\hline  
		\end{tabular}
	\end{center}
\caption{O7 ensemble ($a=0.050$\,fm, $m_\pi=261$\,MeV): Vector form factors at all $Q^2$ values for all extraction methods. \label{tab:Q2_O7}}
\end{table*}


\clearpage


\end{document}